\documentclass[prd,aps,amsfonts,superscriptaddress,longbibliography,notitlepage,twocolumn,10pt]{revtex4-1}

\usepackage{graphicx}
\usepackage{xcolor}
\usepackage{rotating}
\usepackage{amsmath,amssymb,graphics,amsthm,isomath}
\usepackage{bm}
\usepackage{bbold}

\usepackage[colorlinks=true, urlcolor=violet, linkcolor=blue, citecolor=red, hyperindex=true, linktocpage=true]{hyperref}
\usepackage[capitalise,compress]{cleveref}

\makeatletter
\renewcommand{\p@subsection}{}
\renewcommand{\p@subsubsection}{}
\makeatother

\usepackage{xcolor}
\usepackage{mathtools}

\newcommand{\cc}{\mathrm{c}}
\newcommand{\ii}{\mathrm{i}}
\newcommand{\ee}{\mathrm{e}}

\newcommand{\dd}{\mathrm{d}}
\newcommand{\DD}{\mathrm{D}}

\newcommand{\xx}{{\bm{x}}}
\newcommand{\kk}{{\bm{k}}}
\newcommand{\qq}{{\bm{q}}}
\newcommand{\vv}{\bm{v}}
\newcommand{\EE}{\mathcal{E}}
\newcommand{\OO}{\mathcal{O}}
\newcommand{\bG}{\bar{G}}
\newcommand{\intq}{\int\frac{\dd^3 \qq}{(2\pi)^2}}
\newcommand{\dF}{\delta F}

\newcommand{\tq}{{\tilde{q}}}
\newcommand{\tk}{{\tilde{k}}}
\newcommand{\tqq}{{\tilde{\qq}}}
\newcommand{\tkk}{{\tilde{\kk}}}
\newcommand{\tE}{\tilde{\EE}}

\newcommand{\lr}[1]{\left( #1\right)}
\newcommand{\mlr}[1]{\left[ #1\right]}

\newcommand{\alr}[1]{\left\langle #1\right\rangle}
\newcommand{\ilr}[1]{\left| #1\right|}

\newcommand{\tr}[1]{\mathrm{Tr}\lr{#1}}

\newcommand{\ddd}[2]{\frac{\mathrm{d} #1}{\mathrm{d} #2}}

\newcommand{\comment}[1]{}
\usepackage{dsfont}

\def\ba{\begin{eqnarray}}
\def\ea{\end{eqnarray}}

\begin{document}
\title{Information Scrambling of the Dilute Bose Gas at Low Temperature}

\author{Chao Yin}
\email{chao.yin@colorado.edu}
\affiliation{Department of Physics and Center for Theory of Quantum Matter, University of Colorado, Boulder, CO 80309, USA}

\author{Yu Chen}
\email{ychen@gscaep.ac.cn}
\affiliation{Graduate School of China Academy of Engineering Physics, Beijing, 100193, China}

\begin{abstract}
We calculate the quantum Lyapunov exponent $\lambda_L$ and butterfly velocity $v_B$ in the dilute Bose gas at temperature $T$ deep in the Bose-Einstein condensation phase. The generalized Boltzmann equation approach is used for calculating out-of-time ordered correlators, from which $\lambda_L$ and $v_B$ are extracted. At very low temperature where elementary excitations are phonon-like, we find $\lambda_L\propto T^5$ and $v_B\sim c$, the sound velocity. At relatively high temperature, we have $\lambda_L\propto T$ and $v_B\sim c(T/T_*)^{0.23}$. We find $\lambda_L$ is always comparable to the damping rate of a quasiparticle, whose energy depends suitably on $T$. The chaos diffusion constant $D_L=v_B^2/\lambda_L$, on the other hand, differs from the energy diffusion constant $D_E$. We find $D_E\ll D_L$ at very low temperature and $D_E\gg D_L$ otherwise.
\end{abstract}

\date{\today}

\maketitle

\section{Introduction}

Butterfly effect, a defining feature for classical chaotic dynamics, also emerges in quantum settings and is crucial for understanding strongly correlated systems. To diagnose quantum chaos, out-of-time-ordered correlator (OTOC) is first introduced by Larkin and Ovchinikov to study disordered superconductors \cite{LO69}.  This idea is rarely visited until Kitaev recently revived it to understand the shock wave back action in the black hole scattering problem \cite{Kitaev14,Shenker14}. To be specific, we define OTOC by two operators $\OO,\tilde{\OO}$ as
\begin{equation}\label{eq:OTOC}
    {\cal C}(t)={\rm tr}\left( \sqrt{\rho} [{\cal{O} }(t),\tilde{\cal{O}}(0)]^\dagger \sqrt{\rho} [{\cal O}(t),\tilde{\cal O}(0)] \right).
\end{equation}
Here $\rho=Z_\beta^{-1}\ee^{-\beta H}$ with $\beta=1/T$ as the inverse temperature, where we have set the Boltzmann constant $k_B=1$, and $Z_\beta=\tr{\ee^{-\beta H}}$ as the partition function. $H$ is the system Hamiltonian that evolves operators by ${\cal O}(t)=\ee^{\ii tH/\hbar}\OO \ee^{-\ii tH/\hbar}$. For typical chaotic systems, OTOC grows exponentially as $\mathcal{C}(t)\sim c_0\exp(\lambda_L t)$, with $c_0$ being a non-universal constant. $\lambda_L$ is the quantum Lyapunov exponent that measures the growth rate of quantum chaos, which shares similarities and differences with its classical counterpart \cite{Galitski17,Xu20,Yin_top}. It was found that $\lambda_L$ is upper bounded by $2\pi/\beta$ \cite{Maldacena16a}, and the maximal value is saturated by models with gravity duals \cite{Shenker14,Shenker14b, Shenker15,Susskind15}, including the Sachdev-Ye-Kitaev model \cite{Kitaev15,Maldacena16b} dual to Jackiw-Teitelboim gravity \cite{Maldacena16c,Jackiw85,Teitelboim83}. Therefore, calculating $\lambda_L$ is crucial for identifying holographic models \cite{fast_scram13,fast_scram19,Yin_all2all}.

More generally, an information interpretation has been discovered for OTOC \cite{Yoshida16}. Namely, $\lambda_L$ measures how fast local information scrambles to global ones, which reveals the thermalization process in a closed quantum system. Moreover, for systems with a spatial structure, if we define ${\cal O}$ and $\tilde{\cal O}$ as local operators whose locations are of distance $r$, then the OTOC is vanishingly small unless $t\gtrsim r/v_B$, for some constant $v_B$ called the butterfly velocity \cite{Swingle16,Blake_prl,Lucas16}. $v_B$ can be viewed as a $\rho$-dependent extension \cite{Han_vB,Yin_boson} of the Lieb-Robinson velocity \cite{Lieb1972}, the maximal speed information can propagate through the system. Combining $\lambda_L$ with $v_B$, one can define the chaos diffusion constant $D_L=v_B^2/\lambda_L$. In the most chaotic systems, $D_L$ is argued to be universally comparable with charge \cite{Blake_prl,Blake_prd} and energy \cite{E_diff} diffusion constants.

Due to the above implications, general properties of OTOC have arisen a lot of interest (see \cite{Swingle_rev22} for a recent review). For example, OTOC has been theoretically calculated in many-body-localized systems \cite{ChenX16,Fan16,Yu16,Swingle16b,He16}, integrable systems \cite{Luttinger17}, and diffusive metals \cite{Bohrdt17,Swingle17,Liao18}, and experimentally measured in NMR systems \cite{Du17,pai_prl,Li20}, ion traps \cite{otoc_bolinger,IonTrap20} and superconducting circuits \cite{Yao21,Google21,otoc_sc22}. 
However, OTOC remains to be studied for the dilute Bose gas in  Bose-Einstein condensation (BEC), realizable in cold atom experiments \cite{BEC95}. Moreover, unlike models studied before, BEC hosts two temperature regimes with qualitatively different elementary excitations. How does information scramble in the crossover temperature regime? In this paper, we fill this gap using the generalized Boltzmann equations (GBE) approach \cite{Aleiner16,graphene_otoc,gBE_PRE,Pengfei19}.

The rest of the paper is structured as follows. In Section \ref{sec:model}, our model is introduced, where we focus on the BEC regime $T\ll T_{\mathrm{BEC}}$. We identify a crossover temperature $T_*\ll T_{\mathrm{BEC}}$, where quasiparticle excitations change from phonon-like at $T\ll T_*$ to particle-like at $T\gg T_*$. In Section \ref{sec:keld}, we apply the augmented Keldysh formalism to derive GBE that govern the evolution of OTOC, to the leading nontrivial order of the interaction strength $g$. In Section \ref{sec:lambda}, we extract $\lambda_L$ from GBE for the whole temperature regime $T\ll T_{\mathrm{BEC}}$, and get $\lambda_L\propto T^5$ for $T\ll T_*$ and $\lambda_L\propto T$ for $T\gg T_*$. We further show that $\lambda_L$ is comparable to the damping rate of a quasiparticle at a suitably defined energy, which can be extracted from traditional Boltzmann equations. In Section \ref{sec:vB}, we present our results on $v_B$. It is of the order of the sound velocity $c$ at $T\ll T_*$, and grows as a power law $v_B\sim T^{0.23}$ for $T\gg T_*$. We further show that for both temperature regimes, the chaos diffusion constant $D_L$ and the energy diffusion constant $D_E$ are not related to each other. We finally conclude in Section \ref{sec:conclude}.

\section{Model}\label{sec:model}
Here we introduce our model. Consider $N$ bosons contained in a 3-dimensional box of volume $V=L^3$. Using $\psi(\xx)$ to be the complex field operator that annihilates a boson at space position $\xx$, we study the homogeneous Bose gas with Hamiltonian \begin{equation}
    H_{\mathrm{BG}} = H_K + H_V,
\end{equation}
where the kinetic energy is \begin{equation}
    H_K = \int \dd\xx\, \psi^\dagger(\xx) \lr{-\frac{\hbar^2}{2m}\nabla^2} \psi(\xx),
\end{equation}
with $m$ being the boson mass.
The interaction $H_V$ is given by \begin{equation}
    H_V = \frac{g}{2}\int \dd\xx\, \psi^\dagger(\xx)\psi^\dagger(\xx)\psi(\xx)\psi(\xx),
\end{equation}
where we have assumed the temperature is sufficiently low, so that pairs of bosons feel a delta function pseudopotential \cite{zhai_book} \begin{equation}
    v(\xx-\xx') = \frac{4\pi a_s\hbar^2}{m}\delta(\xx-\xx')\equiv g\delta(\xx-\xx'),
\end{equation}
determined by the $s$-wave scattering length $a_s$ (or equivalently, the interaction strength $g$).
In the momentum space, we define the boson annihilation operator at wave vector $\kk$ by $a_\kk =V^{-1}\int\dd\xx\psi(\xx)\ee^{-\ii \kk\cdot\xx}$. Then $H_{\mathrm{BG}}$ can be rewritten as 
\begin{equation}\label{eq:HBG}
    H_{\mathrm{BG}} = \sum_\kk \epsilon_\kk a_\kk^\dagger a_\kk + \frac{g}{2V} \sum_{\kk_1,\kk_2,\kk_3 } a_{\kk_1}^\dagger a_{\kk_2}^\dagger a_{\kk_3} a_{\kk_1+\kk_2-\kk_3},
\end{equation}
where $\epsilon_\kk =\hbar^2k^2/2m$, $k=|\kk|$, and $\kk$ takes values in $\{2\pi n/L: n\in \mathbb{Z}^3\}$.

There are three independent length scales in this model: the scattering length $a_s$, the inter-particle spacing $n^{-1/3}$ where $n=N/V$, and the thermal wavelength \begin{equation}
    \lambda_T = \sqrt{\frac{2\pi\hbar^2}{m T}}.
\end{equation}
We focus on the dilute and low-temperature limit \begin{equation}
    na_s^3 \ll 1,\quad n \lambda_T^3 \gg 1,
\end{equation}
where perturbation theory applies.
In this regime close to equilibrium, nearly all of the $N$ bosons condense in the zero-momentum state, forming a BEC \cite{zhai_book}. As in standard Bogoliubov theory for a homogeneous BEC, we approximate the zero-momentum creation/annihilation operators in \eqref{eq:HBG} by a large $c$-number $\sqrt{N_0}\approx \sqrt{N}$: \begin{equation}\label{eq:bogo}
    a_{\bm 0}=a^\dagger_{\bm 0}=\sqrt{N},
\end{equation}
where we have ignored higher order corrections $N-N_0\propto \sqrt{na_s^3}$ \cite{zhai_book}. 
Moreover, we use the standard Bogoliubov transformation to obtain the effective Hamiltonian from \eqref{eq:HBG}
\begin{subequations}\label{eq:H}
\begin{align}
    H &= H_0+H_1, \quad \mathrm{where} \\
    H_0 &= \sum_\kk \EE_\kk\alpha_\kk^\dagger \alpha_\kk, \quad \mathrm{and} \\
    H_1 &= \frac{g}{\sqrt{V}}\sum_{\kk_1,\kk_2} M_{\kk_1, \kk_2} \lr{ \alpha_{\kk_1}^\dagger \alpha_{\kk_2}^\dagger \alpha_{\kk_1+\kk_2}+\mathrm{h.c.}  }, \label{eq:H1}
\end{align}
\end{subequations}
where $\alpha_\kk$ and $\alpha_\kk^\dagger$ are the annihilation and creation operators for the Bogoliubov quasiparticle, with boson commutation relation \begin{equation}
    [\alpha_\kk, \alpha^\dagger_{\kk'}] = \delta_{\kk,\kk'}.
\end{equation} In \eqref{eq:H}, the quasiparticle has spectrum
\begin{align}\label{eq:EE}
    \EE_\kk = \sqrt{\epsilon_\kk (\epsilon_\kk + 2gn) },
\end{align}
and collision matrix \cite{Matrix01} \begin{equation}\label{eq:Mkk}
    M_{\kk_1, \kk_2} = \sqrt{n} \frac{\mathsf{E}_1+\mathsf{E}_2-\mathsf{E}_3+3\mathsf{E}_1\mathsf{E}_2\mathsf{E}_3}{4\sqrt{\mathsf{E}_1\mathsf{E}_2\mathsf{E}_3}},
\end{equation}
with $\mathsf{E}_i \equiv \epsilon_{\kk_i}/\EE_{\kk_i}$ and $\kk_3 =\kk_1+\kk_2$. In deriving \eqref{eq:H}, we have discarded a $c$-number term, and higher order terms in $1/N$. We have also discarded the term $\propto \alpha_{\kk_1} \alpha_{\kk_2} \alpha_{-\kk_1-\kk_2}+\mathrm{h.c.}$, which describes the process that creates or annihilates three quasiparticles simultaneously. At leading order, such off-shell processes do not contribute to the kinetic equations that we will derive. \eqref{eq:H} is then our starting point for a field-theoretic calculation for information scrambling, and in the end of Section \ref{sec:keld} we will justify the Bogoliubov approximation \eqref{eq:bogo} in this nonequilibrium context. Note that, although strictly speaking, the sums over $\kk$ in \eqref{eq:H} should avoid the $\kk=\bm{0}$ point, this makes no difference for latter calculations, since $\EE_\kk$ and $M_{\kk_1, \kk_2}$ both become zero when one of the $\kk$ arguments (including $\kk_3$) is set to $\bm 0$.

\eqref{eq:EE} suggests a crossover behavior for the quasiparticles. Defining the characteristic momentum $k_0\equiv \sqrt{mgn}/\hbar = \sqrt{4\pi a_s n}$, the quasiparticles change from phonon-like $\EE_\kk \approx \hbar ck$ at $k\ll k_0$, where the sound velocity $c=\sqrt{gn/m}$, to particle-like $\EE_\kk \approx \epsilon_\kk$ at $k\gg k_0$. The corresponding crossover temperature is $T_*\equiv \hbar^2k_0^2/m=\hbar c k_0 = gn$. Thus we expect OTOC also behaves differently at the two temperature regimes: the very low temperature $T\ll T_*$, and the relatively high temperature $T_*\ll T\ll T_{\mathrm{BEC}}$.

\section{The augmented Keldysh Formalism}\label{sec:keld}

\begin{figure}[t]
\centering
\includegraphics[width=0.45\textwidth]{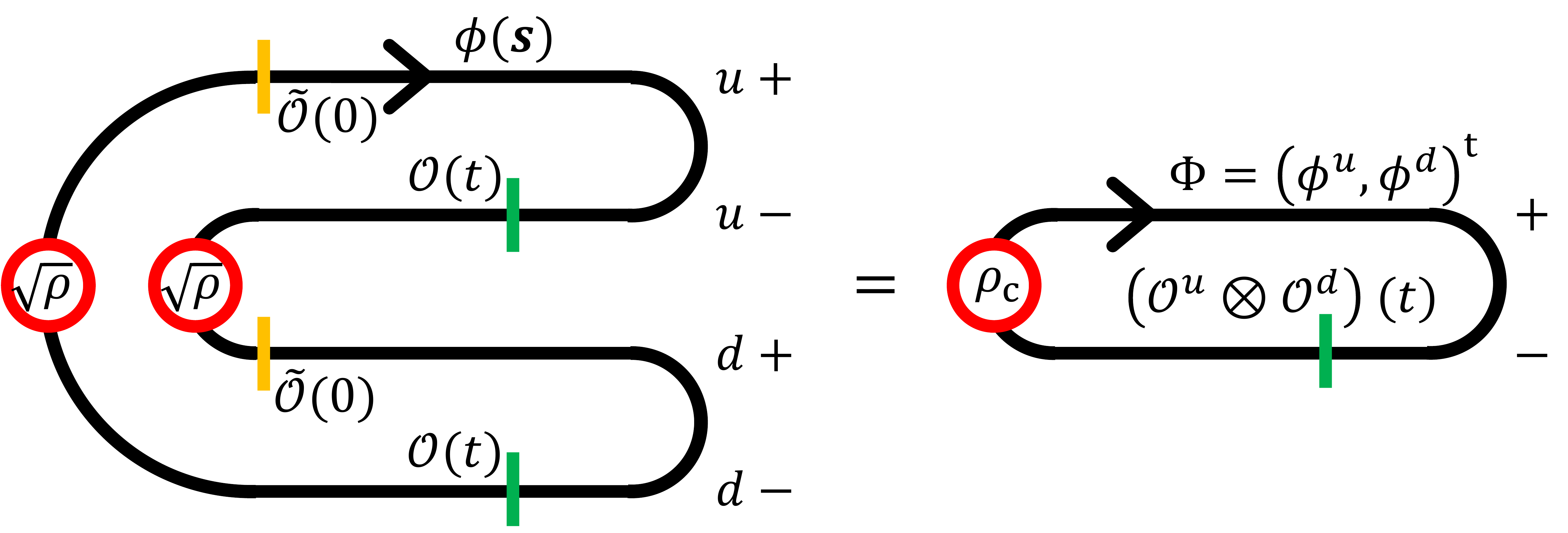}
\caption{\label{fig:contour} The augmented Keldysh contour $C$ (left) for OTOC in \eqref{eq:OTOC}, is equivalent to the conventional Keldysh contour $C_\cc$ (right), where the fields are doubled, and the initial state $\rho_\cc$ includes the perturbation from $\tilde{\OO}$. }
\end{figure}

In this section we set $\hbar=1$. { We first remark on our regularization in \eqref{eq:OTOC}, namely inserting two $\sqrt{\rho}$s between the commutators. The advantage is threefold: It avoids potential ultraviolet divergences, and is the one for which the chaos bound \cite{Maldacena16a} is proved. Moreover, in kinetic theory it has a clear physical meaning related to classical chaos \cite{regularization19}. 

\eqref{eq:OTOC} contains four terms that can be arranged as \begin{align}
    \mathcal{C}(t) &= 2\, \mathrm{Re}\, \tilde{\mathcal{C}}(t) + \text{TOC}, \quad \mathrm{where} \nonumber\\
    \tilde{\mathcal{C}}(t) &= {\rm tr}\left( \sqrt{\rho} {\cal{O} }(t)\tilde{\cal{O}}(0) \sqrt{\rho} {\cal O}(t)\tilde{\cal O}(0) \right). \label{eq:tC}
\end{align}
Here TOC stands for time-ordered correlations, and we have assumed the operators to be Hermitian for simplicity. We focus on $\tilde{\mathcal{C}}(t)$ because TOC does not host exponential growth.

\subsection{relation between OTOC and TOC in a doubled system}
}

To calculate OTOC in \eqref{eq:tC}, we first introduce the time contour $C$ shown on the left of Fig.~\ref{fig:contour}, which contains two \emph{parts}: up($u$) and down($d$), with each part containing two \emph{branches}: for example $u$ contains $u+$ and $u-$. Such $C$ is called the augmented Keldysh contour introduced in \cite{Aleiner16}: if there is only one part (up or down) instead, then it is the conventional Keldysh contour \cite{kamenev_book} that is used for calculating TOC. We parametrize $C$ by the contour time $\bm s$, which goes from $t=0^-$ (the time slightly before $0$) to $t=+\infty$ and back to $t=0^-$ in the up part of $C$, and then goes to $+\infty$ and back to $0^-$ again in the down part of $C$, completing one cycle of the whole contour. Equivalently one can describe the contour time by the doublet ${\bm s}=(\kappa,t)$, where the Keldysh label $\kappa=u+,u-,d+,d-$ denotes the branch that the conventional time $t\in (0^-,+\infty)$ lives in.  Define the contour Hamiltonian $H(\bm s)$ { \begin{equation}
    H(\bm s) = \left\{\begin{array}{cc}
        H -\frac{\ii}{2}\beta H \delta(t+0)& \kappa= u+,d+ \\
        -H & \kappa=u-,d- \\
    \end{array}\right.,
\end{equation}}
where the delta function at $(u+, 0^-)$ and $(d+,0^-)$  accounts for the thermal density matrix $\rho$. Then \eqref{eq:tC} can be rewritten on this contour $C$: \begin{align}\label{eq:contour}
    \tilde{\mathcal{C}}(t) &= \alr{\mathcal{T}_C\OO_{d-}(t) \tilde{\OO}_{d+}(0)\OO_{u-}(t) \tilde{\OO}_{u+}(0) \ee^{-\ii\int_C \dd {\bm s} H(\bm s)} } \nonumber\\ &= \int [\DD \phi]\OO_{d-}(t)\tilde{\OO}_{d+}(0)\OO_{u-}(t) \tilde{\OO}_{u+}(0) \ee^{\ii S[\phi]},
\end{align}
where in the first line, $\mathcal{T}_{C}$ time orders the operators by its position in the contour $C$, and $\alr{\cdot}=Z_\beta^{-1}\tr{\cdot}$. In the second line we used the path integral representation by replacing operators $\alpha_\kk$ and $\alpha_\kk^\dagger$ with classical fields $\phi_\kk(\bm s)$ and $\bar{\phi}_\kk(\bm s)$ that live on the contour $C$, and defined the contour action \begin{equation}
    S[\phi] = S_0[\phi] +S_1[\phi],
\end{equation}
where $S_0$ and $S_1$ correspond to $H_0$ and $H_1$ in \eqref{eq:H} respectively, whose expressions are given later.
We provide several remarks on \eqref{eq:contour}. First, the Keldysh $\kappa$ labels are not unique because the operator insertions can move along the contour: $\OO_{u-}(t)$ can be replaced by $\OO_{u+}(t)$ for example. Second, for notational simplicity we omit the functional dependence of $S$ on $\bar{\phi}$, which is also integrated in $\int[\DD\phi]$. Lastly, we use $\OO_{\kappa}(t)$ for both the quantum operator $\OO$ at ${\bm s}=(\kappa,t)$, and its path integral representation that is a function of $\phi(\bm s)$ and its time derivatives.

We have expressed \eqref{eq:tC} as a path integral along the augmented Keldysh contour $C$, which gets rid of operators and their time ordering. As a result, there is an equivalent perspective that turns out to be useful: The path integral can be viewed as one along a \emph{conventional} Keldysh contour $C_\cc$ as shown on the right of Fig.~\ref{fig:contour} instead, by merging the up and down parts of $C$, so that there are two sets of fields $\phi^u(\bm s)$ and $\phi^d(\bm s)$ that live on the contour $C_\cc$. Here $\bm s$ is the contour time for $C_\cc$, and we combine the fields to a two-component one $\Phi=(\phi^u, \phi^d)^{\mathrm{t}}$ with its conjugate $\bar{\Phi}=(\bar{\phi}^u, \bar{\phi}^d)$. 
{ The operator insertions are also combined, where the initial perturbations $\tilde{\OO}$ are absorbed into the initial state $\rho_\cc$. Later we will find the specific form of $\OO,\tilde{\OO}$ and $\rho_\cc$ is irrelevant for us to extract $\lambda_L$ and $v_B$. } The action governing the contour evolution in $0<t<\infty$ factorizes to up and down contributions, so that the OTOC is converted to a TOC $\alr{\OO_{u-}(t)\OO_{d-}(t)}$, for a doubled system: the original one, $u$, together with its augmented ancilla system $d$. 
{ Here we call $\alr{\OO_{u-}(t)\OO_{d-}(t)}$ a TOC because it can be calculated on a single Keldysh contour. To be more precise, it can viewed as $\alr{(\OO\otimes\OO)(t)I(0)}$, where the two $\OO$s are combined to one operator $\OO\otimes \OO$, and an identity operator is inserted at time $0$ to make the time order manifest. } The two subsystems have the same Hamiltonian \eqref{eq:H} for time evolution, and do not couple to each other. However, there is a price to pay: The initial state $\rho_\cc$, for the average $\alr{\cdot}$ appearing in the TOC, 
{ includes the perturbation $\tilde{\OO}$ and} becomes a complicated entangled state shared by the two subsystems, which is expressed pictorially in Fig.~\ref{fig:contour}. (Without the perturbation, the density matrix for each subsystem is the exact thermal state $\rho$, 
{ because tracing $d$, for example, is equivalent to removing the two operators $\tilde{\OO}(0),\OO(t)$ in the $d$ part of the left of Fig.~\ref{fig:contour}, so that the two $\sqrt{\rho}$ insertions combine to one $\rho$ as the initial state of $u$}.) 
%\yu{Considering to replace the above sentence by: Without the perturbation, one can find the OTOC is the Wightman Green function\cite{Aleiner16}. }
This perturbed initial entangled state leads to correlations shared by the two subsystems, and the growth of $\mathcal{C}(t)$ measures how such correlations, probed by the local operator $\OO$, decay when evolving from the initial state $\rho_\cc$. At long times $\mathcal{C}(t)$ stays at some large value, which means the two subsystems have \emph{locally} forgotten about the initial condition and become uncorrelated \cite{Aleiner16}. 

{ 
\subsection{overview of the derivation}
With the above relation to TOC in the doubled system $u+d$, it is transparent that conventional Keldysh techniques (see \cite{kamenev_book} for a pedagogical review) apply here with slight modifications. Here we sketch the idea before diving into technical details. 

Without interaction, the problem is solvable by explicit single-particle Green function $G_0$, which contains three exponents: ``retarded'' $G^R_0$, ``advanced'' $G^A_0$ and ``Keldysh'' $G^K_0$. With interaction, the full Green function $G$ is related to $G_0$ via the self-energy $\Sigma$ in the Dyson equation \eqref{eq:dyson}, and some approximation needs to made.

First, we take the semi-classical approximation so that the Dyson equation for $G^K$ amounts to a kinetic Boltzmann equation for some quasiparticle distribution function $F(t,\bm{x}, \bm{k})$, whose initial value is determined by $\rho_\cc$. This requires that the initial state $\rho_\cc$, perturbed by $\tilde{\OO}$, fluctuates in length scales much larger than the microscopic ones. 

Second, since the interaction is weak, we take the self-consistent Born approximation \cite{kamenev_book} for $\Sigma$, namely setting $G^R,G^A$ to their non-interacting counterparts while keeping the full $G^K$ expressed by $F$. This leads to \emph{nonlinear} partial differential equations for $F$, the GBEs. It is argued that considering further contributions beyond this approximation does not change the form of the resulting GBE \cite{Aleiner16}, because it merely changes the spectrum and interaction vertex in a non-qualitative way.

Thirdly, we linearize the GBE assuming there is a time window in which $F$ is close to its unstable fixed point $F_0$ of the GBE, which turns out to be the value without interaction and operator perturbation. Since the expectation of a general local operator $\OO$ is a function of the distribution $F$, we find $\lambda_L,v_B$ simply by extracting the fastest growing mode $F-F_0\sim \ee^{\lambda_L(t-x/v_B)}$ from the GBE on how $F$ deviates from $F_0$. The result then does not depend on the specific form of $\OO$, and the initial state $\rho_\cc$ that includes the interacting density matrix $\rho$ and $\tilde{\OO}$. We only require that $\rho$ is close to the non-interacting $\rho_0$, and that $\tilde{\OO}$ is weakly perturbing and ``smeared out'' (justifying our first approximation above). As another perspective, $\rho$ becomes irrelevant by arguing that it can be viewed as the state evolved from the non-interacting $\rho_0$ in the far past $t=-\infty$, with interaction adiabatically turned on \cite{kamenev_book}.
}

%\yu{Thirdly, when $\tilde{\cal O}$ is taken to be unity operator, the OTOC we defined is nothing but Wightman Green fucntion at thermal equilibrium. Then the generalized Boltzmann equation will give a distribution function in up-down sector related to Wightman Green function. One see also find that the Dyson equation for the doubled system works for general $\rho_c$. Then we suppose $\rho_c=\rho_{\rm thermal}+\delta\rho$ is a perturbation from thermal distribution, and it translates to a small deviation in distribution function $F=F_{\rm thermal}+\delta F$. Then we find $\lambda_L$ can be extracted from generalized Boltzmann equation for $\delta F$ in a form of $\partial_t \delta F={\cal M}\delta F$. $\lambda_L$ is the maximal eigen value of matrix ${\cal M}$, and it describes how fast the perturbation can send the system out-of-equilibirum. Initially, we assume of homogenous $\delta F$, then we obtain only information for $\lambda_L$. If we assume $\delta F$ is position dependent, then we can extract chaos diffusion rate $v_B$ from the linearized extended Boltzmann equation. }

\subsection{Keldysh rotation}
We first focus on the noninteracting case in this subsection to motivate such techniques, which also provides building blocks for the interacting case. The noninteracting action on contour $C_\cc$ is \begin{equation}
    S_0[\Phi] = \int \dd t \sum_{s=\pm}\sum_\kk s\, \bar{\Phi}_\kk^{s}(t)\lr{\ii \partial_t-\EE_\kk } \Phi_\kk^{s}(t),
\end{equation}
where the kernel $\ii \partial_t-\EE_\kk $ should be understood as a diagonal matrix acting on the $(u,d)$ space. $s=\pm$ is the branch index, $+$ for forward time evolution and $-$ for backwards.
Despite of the factorized form of $S_0$, the two sets of fields $\Phi^+$ and $\Phi^-$ are correlated because the two branches are connected at $t=0$ and $t=\infty$. The connection at $t=\infty$ is a trivial continuity condition, while that at $t=0$ involves inserting the initial state $\rho_\cc$. Due to these connections, the Green functions $\alr{\Phi^{s}(t)\bar{\Phi}^{s'}(t')}_0$ satisfy exact causality conditions. For example, for $t'>t$ we have $\alr{\Phi(t)^{+}\bar{\Phi}^+(t')}_0 = \alr{\Phi(t)^{+}\bar{\Phi}^-(t')}_0$ by moving $\bar{\Phi}^+$ from $t'$ along the contour to $\infty$ and then back to $t'$, with the field becoming $\bar{\Phi}^-$. To make such conditions manifest, we pursue the Keldysh rotation \cite{kamenev_book}: \begin{equation}
    \lr{\begin{array}{c}
    \Phi^1(t)  \\
    \Phi^2(t)
    \end{array}} = \frac{1}{\sqrt{2}}\lr{\begin{array}{cc}
    1 & 1  \\
    1 & -1
    \end{array}} \lr{\begin{array}{c}
    \Phi^+(t)  \\
    \Phi^-(t)  
    \end{array}},
\end{equation}
where $\Phi^1$ and $\Phi^2$ are often referred to as the ``classical'' and ``quantum'' field respectively.
The new fields have Green functions of the form \begin{equation}\label{eq:GKRA}
    \alr{\Phi^s(t) \bar{\Phi}^{s'}(t')}_0 \equiv \ii G_0^{ss'}(t,t') = \ii\lr{\begin{array}{cc}
    G_0^K(t,t') & G_0^R(t,t')  \\
    G_0^A(t,t') & 0
    \end{array}},
\end{equation}
where $K,R,A$ stand for ``Keldysh'', ``retarded'' and ``advanced'', and the zero matrix element is due to causality. Here index $s=1,2$ is introduced to label the degrees of freedoms in the retard/advanced (RA) space.  Recall that $G_0^K, G_0^R, G_0^A$ are themselves matrices in the $(u,d)$ space: \begin{equation}\label{eq:GR=GR}
    G_0^K = \lr{\begin{array}{cc}
    G_0^{Kuu} & G_0^{Kud} \\
    G_0^{Kdu} & G_0^{Kdd}
    \end{array}},\quad\!\!\! G_0^{R/A} = \lr{\begin{array}{cc}
    G_0^{R/Auu} & 0 \\
    0 & G_0^{R/Add}
    \end{array}}.
\end{equation}
Thus the Green functions can be labeled as $G^{\kappa\kappa'}=G^{ss',\sigma\sigma'}$, 
where $\sigma=u,d$ is introduced for the up/down index. One can also notice that $G^{R/A uu}=G^{R/A dd}$ due to causality \cite{Aleiner16}.
In the absence of the initial perturbation $\tilde{\OO}$, the system is in equilibrium so that the Green functions only depend on the time difference $\bG^{\kappa\kappa'}_0(t,t')=\bG^{\kappa\kappa'}_0(t-t')$, with the symbol $\bar{\cdot}$ denoting equilibrium. Then $\bG_0$ can be Fourier transformed to frequency space $\bG^{\kappa\kappa'}_0(\omega) = \int \dd t \bG^{\kappa\kappa'}_0(t)\ee^{\ii\omega t}$. From the specific form of $\rho_\cc$ without the perturbation, one can derive \cite{kamenev_book,Aleiner16} \begin{subequations} \label{eq:G0} \begin{align}
    &G_{0,\kk}^{R/A uu}(\omega) =G_{0,\kk}^{R/A dd}(\omega)= G_0^{R/A}(\omega)= \lr{\omega-\EE_\kk\pm\ii 0}^{-1}, \label{eq:G0R} \\
    &\bG_{0,\kk}^{Kuu}(\omega) = \bG_{0,\kk}^{Kdd}(\omega) = -2\pi\ii \coth\lr{\frac{\omega}{2T}}\delta(\omega-\EE_\kk), \\
    &\bG_{0,\kk}^{Kdu}(\omega) = \bG_{0,\kk}^{Kud}(\omega) = -2\pi\ii \lr{\sinh\lr{\frac{\omega}{2T}}}^{-1}\delta(\omega-\EE_\kk).
\end{align}
Here we use $G_0^{R/A}$ instead of $\bG_0^{R/A}$, because of the noninteracting nature that retarded/advanced Green functions do not depend on the initial state \cite{kamenev_book}: For example, \eqref{eq:G0R} holds even when the perturbation $\tilde{\OO}$ is present.
\end{subequations} 
The up/down diagonal elements of $\bG_0$ agree with the conventional Keldysh result, since the initial density matrix for each subsystem is just $\rho$. In particular, these Green functions satisfy the fluctuation-dissipation theorem (FDT) \begin{align}\label{eq:FDT}
    \bG_0^{Kuu}(\omega) &= F_0(\omega)\mlr{G_0^R(\omega) -G_0^A(\omega)},\quad \mathrm{where} \nonumber\\ F_0(\omega) &= \coth\lr{\frac{\omega}{2T}}.
\end{align}
Similarly, one can write down the generalized version of FDT for the off-diagonal element, where the two subsystems are jointly probed by the fields: \begin{align}\label{eq:FDT1}
    \bG_0^{Kdu}(\omega) &= F_0^{du}(\omega)\mlr{G_0^R(\omega) -G_0^A(\omega)},\quad\mathrm{where} \nonumber\\ F_0^{du}(\omega) &= \lr{\sinh\lr{\frac{\omega}{2T}}}^{-1}.
\end{align}

\subsection{the generalized Boltzmann equations}
\begin{figure}[t]
\centering
\includegraphics[width=0.4\textwidth]{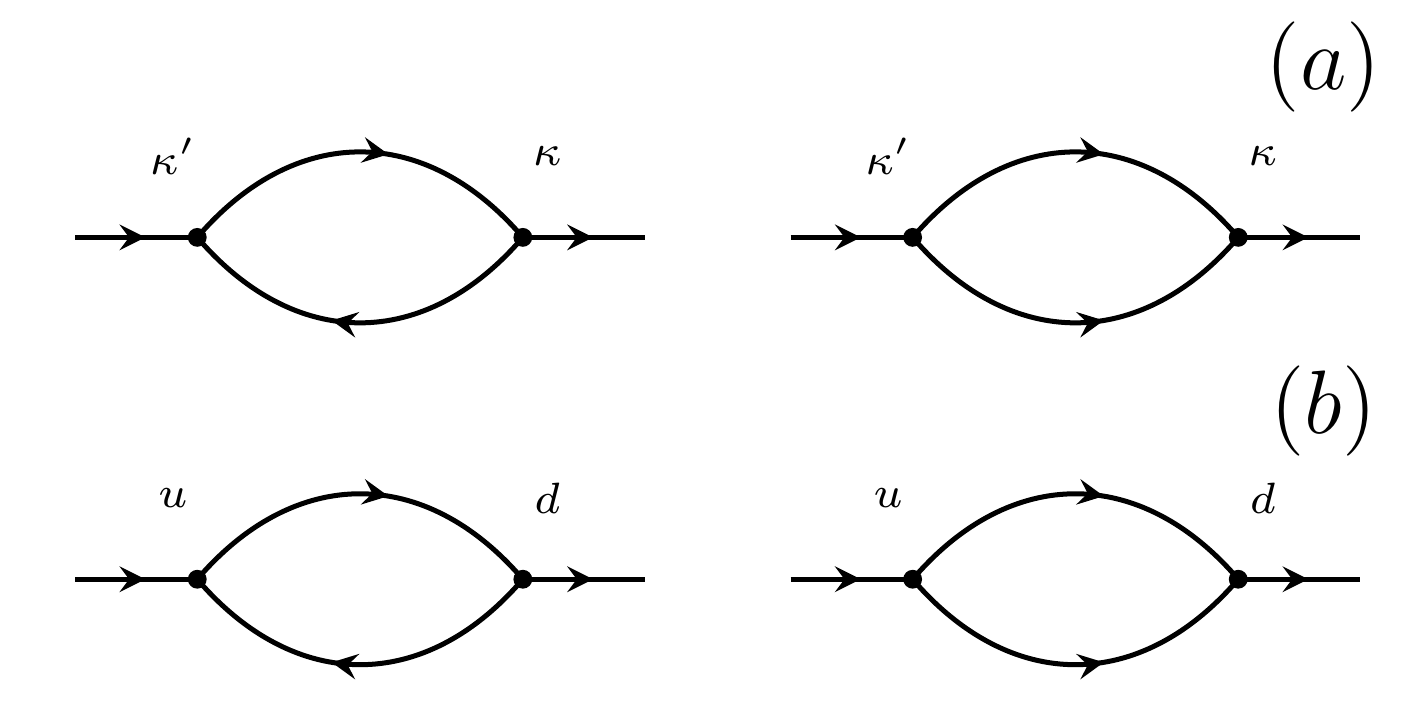}
\caption{\label{fig:sigma} Feynman diagrams for the self-energy (a) $-\ii\Sigma^{\kappa\kappa'}$, and in particular, (b) $-\ii\Sigma^{Kdu}$. (a) The Keldysh labels $\kappa,\kappa'$ are in RA and $ud$ space. The first Feynman diagram can be viewed as a virtual Landau damping followed by a Belieav damping. The second diagram can be viewed as a virtual Belieav damping followed by a Landau damping process. (b) $-\ii\Sigma^{Kdu}$ as a special case of (a). The up/down index is denoted at the vertices, since $u$ and $d$ do not mix by the interaction \eqref{eq:S1}. Furthermore, the internal lines only involve $G^{Kud}$ and $G^{Kdu}$ due to the RA structure in \eqref{eq:S1}, and no spin index is summed over.}
\end{figure}

Having formalized the noninteracting theory for $H_0$, we treat $H_1\propto g$ perturbatively and calculate the full Green function $G^{ss'}(t,t') = -\ii\alr{\Phi^s(t) \bar{\Phi}^{s'}(t')}$ to second order of $g$. To this end, we first note that $G$ can be written in the form of \eqref{eq:GKRA} and \eqref{eq:GR=GR}, with all $0$s removed in subscripts, because $G$ obey the same causality conditions as $G_0$. We start with writing down the interaction action from \eqref{eq:H1}, \begin{align}\label{eq:S1}
    S_1[\phi] = \frac{-g}{\sqrt{2V}}\int \dd t \sum_{\sigma=u,d} \sum_{\kk_1,\kk_2} &( \bar{\phi}_1^{\sigma1}\bar{\phi}_2^{\sigma1}\phi_3^{\sigma2} + 2\bar{\phi}_1^{\sigma1}\bar{\phi}_2^{\sigma2}\phi_3^{\sigma1} \nonumber\\ &+ \bar{\phi}_1^{\sigma2}\bar{\phi}_2^{\sigma2}\phi_3^{\sigma1} + \mathrm{c.c.}),
\end{align}
where the Keldysh rotation has been performed, and $\phi_j$ is the shorthand notation for $\phi_{\kk_j}$, with $\kk_3=\kk_1+\kk_2$ being implicit. Expanding the path integral in powers of $g$, we calculate the self-energy \begin{equation}\label{eq:sigma}
    \Sigma = \lr{\begin{array}{cc}
    0 & \Sigma^A  \\
    \Sigma^R & \Sigma^K
    \end{array}},
\end{equation} 
which corresponds to the one-particle irreducible diagrams for the Green function. When expanded to the up/down basis, we have \begin{equation}\label{eq:SR=SR}
    \Sigma^K = \lr{\begin{array}{cc}
    \Sigma^{Kuu} & \Sigma^{Kud}  \\
    \Sigma^{Kdu} & \Sigma^{Kdd}
    \end{array}}, \quad \Sigma^{R/A} = \lr{\begin{array}{cc}
    \Sigma^{R/A} & 0  \\
    0 & \Sigma^{R/A}
    \end{array}},
\end{equation}
whose corresponding Feynman diagrams are summarized in Fig.~\ref{fig:sigma} for the leading order $\sim g^2$.
The self-energy $\Sigma$ is related to the full Green function by the Dyson equation \begin{equation}\label{eq:dyson}
    \lr{\hat{G}_0^{-1}-\hat{\Sigma}}\circ \hat{G} = \hat{\mathbb{1}},
\end{equation}
where the hat symbol $\hat{A}$ means that $A$ is viewed as a matrix acting on the direct product space of the momentum-frequency space (which is suitably discretized), and the four-dimensional augmented Keldysh space. The symbol $\circ$ means the matrix multiplication on this direct product space, which involves the convolution in the continuous space-time. \eqref{eq:dyson} can be rewritten as $\hat{\Sigma} = \hat{G}_0^{-1}-\hat{G}^{-1}$, so that the causality structure of $G$ and $G_0$ gives rise to the structure of $\Sigma$ in \eqref{eq:sigma} and \eqref{eq:SR=SR}.

Motivated by \eqref{eq:FDT} and \eqref{eq:FDT1}, we introduce the Hermitian distribution matrix $\hat{F}$ to encode the initial condition at $t=0$: \begin{equation}\label{eq:GK=F}
    \hat{G}^K = \hat{G}^R \circ \hat{F} - \hat{F}\circ \hat{G}^A,
\end{equation}
Plugging this parametrization into the Dyson equation \eqref{eq:dyson}, we get \begin{equation}\label{eq:GF-FG}
    \lr{\hat{G}^R_0}^{-1} \circ \hat{F} - \hat{F} \circ \lr{\hat{G}^A_0}^{-1} = \hat{\Sigma}^R\circ \hat{F} - \hat{F} \circ\hat{\Sigma}^A - \hat{\Sigma}^K,
\end{equation}
where we have discarded a term $\lr{\hat{G}^K_0}^{-1}$ that is infinitesimal due to \eqref{eq:G0R}.
The kinetic equation \eqref{eq:GF-FG} is formally exact, and determines the evolution of $F$ if $\Sigma$ is expressed as a functional of $F$ in a self-consistent way, as we will show in the next subsection. 

However, \eqref{eq:GF-FG} is difficult to solve in general.
In order to get a semi-classical Boltzmann-like version from \eqref{eq:GF-FG}, we take the standard assumption \cite{kamenev_book} that the dynamics perturbed from equilibrium varies slowly in space and time, compared to the microscopic scales. Then for any two point function such as the distribution function $F(x_1,x_2) = F(\xx_1, t_1, \xx_2, t_2)$, we perform the Wigner transformation \begin{equation}
    F(x,p) = \int \dd x' \ee^{-\ii p x'} F\lr{x+\frac{x'}{2}, x-\frac{x'}{2}},
\end{equation}
where $p=(\omega,\kk)$ and $px'\equiv \kk\cdot \xx' - \omega t'$. Assuming such functions vary slowly with $x$, one can expand the convolution in their derivatives. For example, $\hat{\Sigma}^R\circ \hat{F}$ is Wigner transformed to \begin{align}
    \lr{\Sigma^R F}(x,p)\approx &\Sigma^R(x,p) F(x,p) \nonumber\\ &+ \frac{\ii}{2}\lr{\partial_x \Sigma^R\partial_p F - \partial_p \Sigma^R\partial_x F},
\end{align}
with the arguments $(x,p)$ being implicit in the second line. Furthermore, since to leading order $F(x,p)$ always appear with \begin{equation}\label{eq:GR-A}
    G_0^{Ruu/dd}(p) - G_0^{Auu/dd}(p)=-2\pi\ii \delta(\omega-\EE_\kk),
\end{equation} 
according to \eqref{eq:GK=F}, we can set the argument $\omega$ of $F(x,\kk,\omega)$ on-shell: \begin{equation}\label{eq:onshell}
    F(x,\kk,\EE_\kk)\rightarrow F(x,\kk),
\end{equation}
so that the reduced distribution $F(x,\kk)$ is interpreted semi-classically as the quasiparticle distribution function at time $t$, position $\xx$ and momentum $\kk$. Using the above two approximations, \textit{i.e.}, derivative expansion and on-shell approximation, \eqref{eq:GF-FG} becomes the GBE \begin{equation}\label{eq:boltz}
    \mlr{(Z')^{-1}\partial_t + \vv'_\kk\cdot \nabla_\xx -\lr{\nabla_\xx \mathrm{Re}\Sigma^R}\cdot\nabla_\kk }F = \mathrm{St}[F],
\end{equation}
where \begin{equation}
    (Z')^{-1}=1-\partial_\omega \mathrm{Re}\Sigma^R, \quad \vv'_\kk= \nabla_\kk\lr{\EE_\kk+\mathrm{Re}\Sigma^R},
\end{equation}
and the collision integral \begin{equation}\label{eq:stf}
    \mathrm{St}[F] = \left.\lr{\ii\Sigma^K+2F\, \mathrm{Im}\Sigma^R}\right|_{\omega =\EE_\kk}.
\end{equation}
Here we have used $\Sigma^A=\lr{\Sigma^R}^*$. From now on, we work with leading order of the interaction strength $g$. Then the terms $\propto \mathrm{Re}\Sigma^R$ on the left hand side of \eqref{eq:boltz} can be ignored, since the spatial and time derivatives are already small in $g$ according to the right hand side.

\subsection{self-energy calculation}
In this subsection we express the self-energy $\Sigma(x,p)$ using the distribution function $F(x,p)$, so that \eqref{eq:boltz} becomes a closed dynamical equation of $F$. In the spirit of the derivative expansion above, $\Sigma(x,p)$ only depends on the local $F(x,p')$ at the same space-time $x$, so we ignore the $x$ label below. The Feynman diagrams at leading order $\Sigma \sim g^2$ are shown in Fig.~\ref{fig:sigma}(a), where the Keldysh labels $\kappa,\kappa'$ are viewed as spin indices. The internal propagators involve all three types of Green function in \eqref{eq:GKRA}. We set the retarded/advanced propagators to be the bare $G_0^{R/A}$ in \eqref{eq:G0R}, which do not depend on the initial state. In contrast, we set the Keldysh propagators to be the nonperturbative $G^K$ that depends on $F$ via \eqref{eq:GK=F}, in which $\hat{G}^R$ and $\hat{G}^A$ are again replaced by its bare counterpart. In this way we self-consistently ``resum'' the contributions from the nonequilibrium distribution $F$, while keeping the spectral Green functions $G^{R/A}$ at leading orders. This resummation will lead to \emph{nonlinear} partial differential equations for $F$. 

As an explicit example, we derive the off-diagonal self-energy $\Sigma^{Kdu}$ in detail. The Feynman diagrams for $\Sigma^{Kdu}$ are shown in Fig.~\ref{fig:sigma}(b), which only involve $G^{Kud}$ and $G^{Kdu}$ as internal propagators. For the left diagram in Fig.~\ref{fig:sigma}(b), the two internal lines are $\ii G^{Kdu}(q)$ and $\ii G^{Kud}(p+q)$ with $p=(\omega,\kk)$ and $q=(q_0,\qq)$ being the external and loop four-momentum. The vertices correspond to the second term in the bracket in \eqref{eq:S1} and its complex conjugate, which contribute a factor $\lr{\frac{-2\ii g}{\sqrt{2V}}M_{\kk,\qq}}^2$. Finally, we sum over momentum $\qq$ and integrate over frequency $\int\frac{\dd q_0}{2\pi}$ to get the contribution from the left diagram \begin{align}
    &-\ii\Sigma^{Kdu}_{\mathrm{L}}(p) = 2g^2\int\frac{\dd^4 q}{(2\pi)^4} M^2_{\kk,\qq}G^{Kdu}(q)G^{Kud}(p+q), \nonumber\\ &= -2g^2\intq M^2_{\kk,\qq}F^{du}_\qq F^{du}_{\kk+\qq}\delta(\omega+\EE_\qq-\EE_{\kk+\qq}),
\end{align}
where in the first line we have replaced $V^{-1}\sum_\qq = \int\frac{\dd^3 \qq}{(2\pi)^3}$, and in the second line we have used \eqref{eq:GK=F}, \eqref{eq:GR-A} and \eqref{eq:onshell}. We have also used the fact that $F$ is Hermitian, $F^{ud}=F^{du}$, and the shorthand notation $F_\qq\equiv F(\qq)$. Similarly, the right diagram in Fig.~\ref{fig:sigma}(b) corresponds to \begin{equation}
    \ii\Sigma^{Kdu}_{\mathrm{R}}(p) = g^2\intq M^2_{\qq,\kk-\qq}F^{du}_\qq F^{du}_{\kk-\qq}\delta(\omega-\EE_\qq-\EE_{\kk-\qq}),
\end{equation}
where one needs to take a symmetry factor $2$ into account. The total off-diagonal self-energy is then $\Sigma^{Kdu} = \Sigma^{Kdu}_{\mathrm{L}} + \Sigma^{Kdu}_{\mathrm{R}}$.

Calculating other components of $\Sigma$ in a similar way, we get the collision integral \eqref{eq:stf} at leading order: \begin{widetext}
\begin{align}\label{eq:stud}
    \mathrm{St}^{ud}_\kk=g^2 \intq
    \Big\{& M^2_{\qq, \kk-\qq}\delta(\EE_\kk-\EE_\qq-\EE_{\kk-\qq})\mlr{F^{du}_\qq F^{du}_{\kk-\qq} - \lr{F^{uu}_\qq + F^{uu}_{\kk-\qq}} F^{du}_\kk} \nonumber\\
    +&2M^2_{\kk, \qq}\delta(\EE_\kk+\EE_\qq-\EE_{\kk+\qq})\mlr{F^{du}_\qq F^{du}_{\kk+\qq} - \lr{F^{uu}_\qq - F^{uu}_{\kk+\qq}} F^{du}_\kk}\Big\}. \\
    \mathrm{St}^{uu}_\kk=g^2\int \frac{\dd^3\bm{q}}{(2\pi)^2}
    \Big\{& M^2_{\qq, \kk-\qq}\delta(\EE_\kk-\EE_\qq-\EE_{\kk-\qq})\mlr{F^{uu}_{\bm{q}} F^{uu}_{\kk-\bm{q}}+1-(F^{uu}_{\bm{q}}+ F^{uu}_{\kk-\bm{q}})F^{uu}_{\bm{k}}} \nonumber\\
    +&2M^2_{\kk, \qq}\delta(\EE_\kk+\EE_\qq-\EE_{\kk+\qq})\mlr{F^{uu}_{\bm{q}} F^{uu}_{\kk+\bm{q}}-1-(F^{uu}_{\bm{q}}- F^{uu}_{\kk+\bm{q}})F^{uu}_{\bm{k}}}\Big\}. \label{eq:stuu}
\end{align}
\end{widetext}
On the other hand, $\mathrm{St}^{du}$ and $\mathrm{St}^{dd}$ are simply related by $u\leftrightarrow d$ symmetry. A crucial observation is that the diagonal $\mathrm{St}^{uu}$ is just the collision integral for TOC of subsystem $u$, which does not involve the off-diagonal $F^{ud}$. The reason is the two subsystems $u$ and $d$ evolve independently in time, and they are correlated only from the initial state $\rho_{\mathrm{c}}$. As a consequence, when the operator $\tilde{\OO}$ perturbs the system away from the unperturbed equilibrium \eqref{eq:FDT} and \eqref{eq:FDT1}, $F^{\kappa\kappa'} = F^{\kappa\kappa'}_0 + \dF^{\kappa\kappa'}$ where $F^{uu}_0=F^{dd}_0=F_0$, there are two classes of eigen-modes for $\dF^{\kappa\kappa'}$ as the solutions for the GBE \eqref{eq:boltz}. In the first class, both the diagonal and off-diagonal components of $\dF$ are nonvanishing, and the diagonal ones evolve independently according to \eqref{eq:stuu}. Since the perturbation is from the stable equilibrium \eqref{eq:FDT}, perturbations of this class are generally decaying modes $\dF^{\kappa\kappa'}(t) \propto \ee^{\lambda t}$ with $\lambda<0$, so that the system returns to equilibrium at long time, guaranteed by the Boltzmann H-theorem. In the second class, the diagonal ones vanish $\dF^{uu}=\dF^{dd}=0$, and the off-diagonal $\dF^{du}\sim \ee^{\lambda t}$, where now the $\lambda$ is no longer guaranteed to be negative. If there is some eigen-mode with $\lambda>0$, it dominates at long times when the first class eigen-modes can be ignored. Therefore, to extract the Lyapunov exponent and butterfly velocity, it suffices to focus on the off-diagonal component of the GBE \eqref{eq:boltz}, with collision integral \eqref{eq:stud}, where the diagonal distributions $F^{uu}=F^{dd}=F_0$ are set to equilibrium \eqref{eq:FDT}.

We have established the GBE describing OTOC dynamics for the effective Hamiltonian \eqref{eq:H}, which comes from the original Bose gas Hamiltonian \eqref{eq:HBG} via the Bogoliubov approximation \eqref{eq:bogo}. We now justify this approach in our nonequilibrium context. According to the previous paragraph, we are interested in the time scale long enough so that the two subsystems $u$ and $d$ have been in equilibrium, as probed locally in each subsystem. Similar to $\delta F^{uu}$ that has already decayed at this time scale, whatever perturbations to the condensate of each subsystem caused by $\tilde{\OO}$ have also died out, so that \eqref{eq:bogo} holds.  Note that we also require this time scale is not too long, so that the inter-subsystem probe $\delta F^{du}$ has not grown beyond the linear regime. We also mention that we have discarded off-shell terms in \eqref{eq:H1}. This approximation is also legitimate, because the collision integral, \eqref{eq:stud} for example, involves only on-shell processes at leading order.

\section{Lyapunov exponent}\label{sec:lambda}
\begin{figure}[t]
\centering
\includegraphics[width=0.45\textwidth]{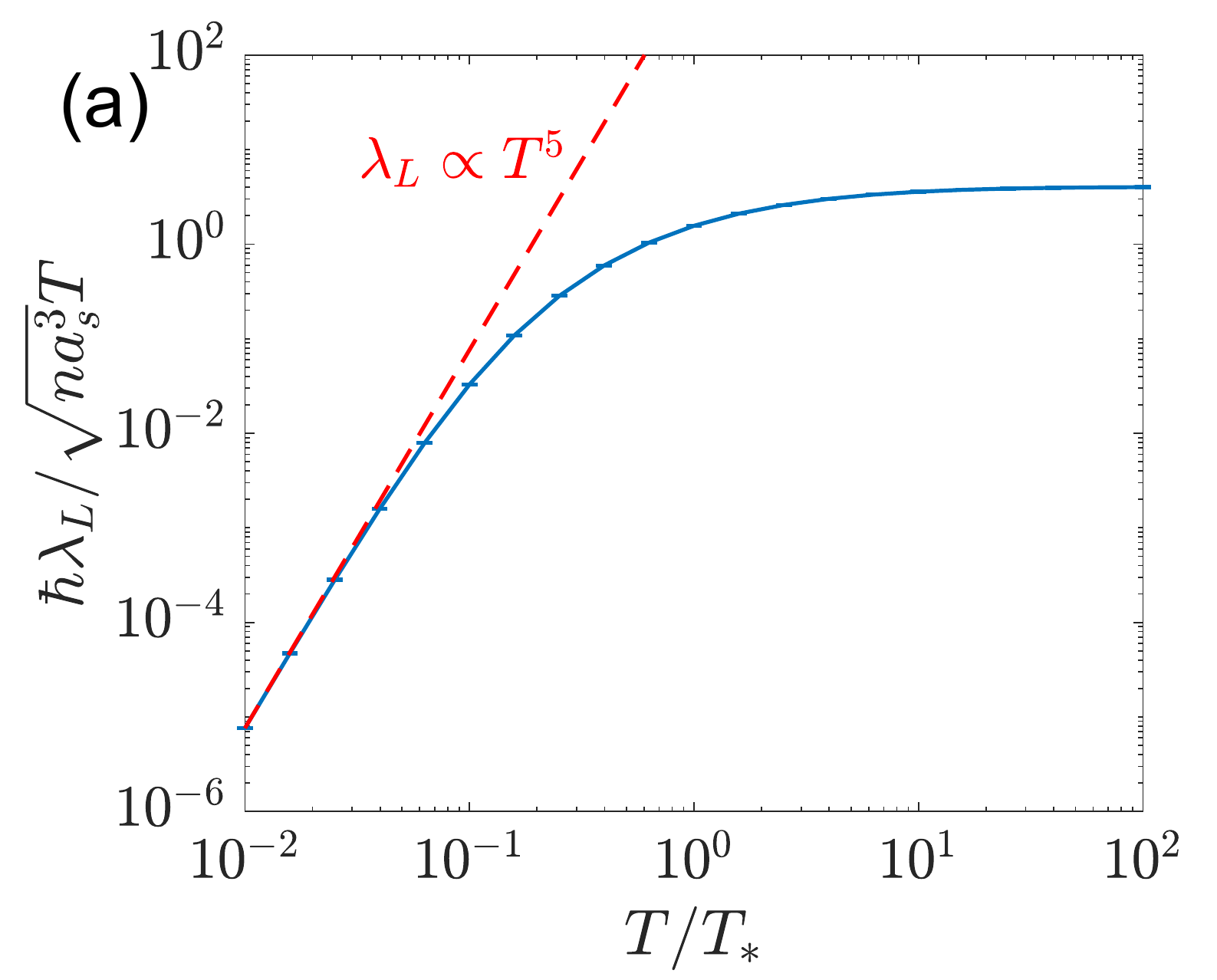}
\includegraphics[width=0.45\textwidth]{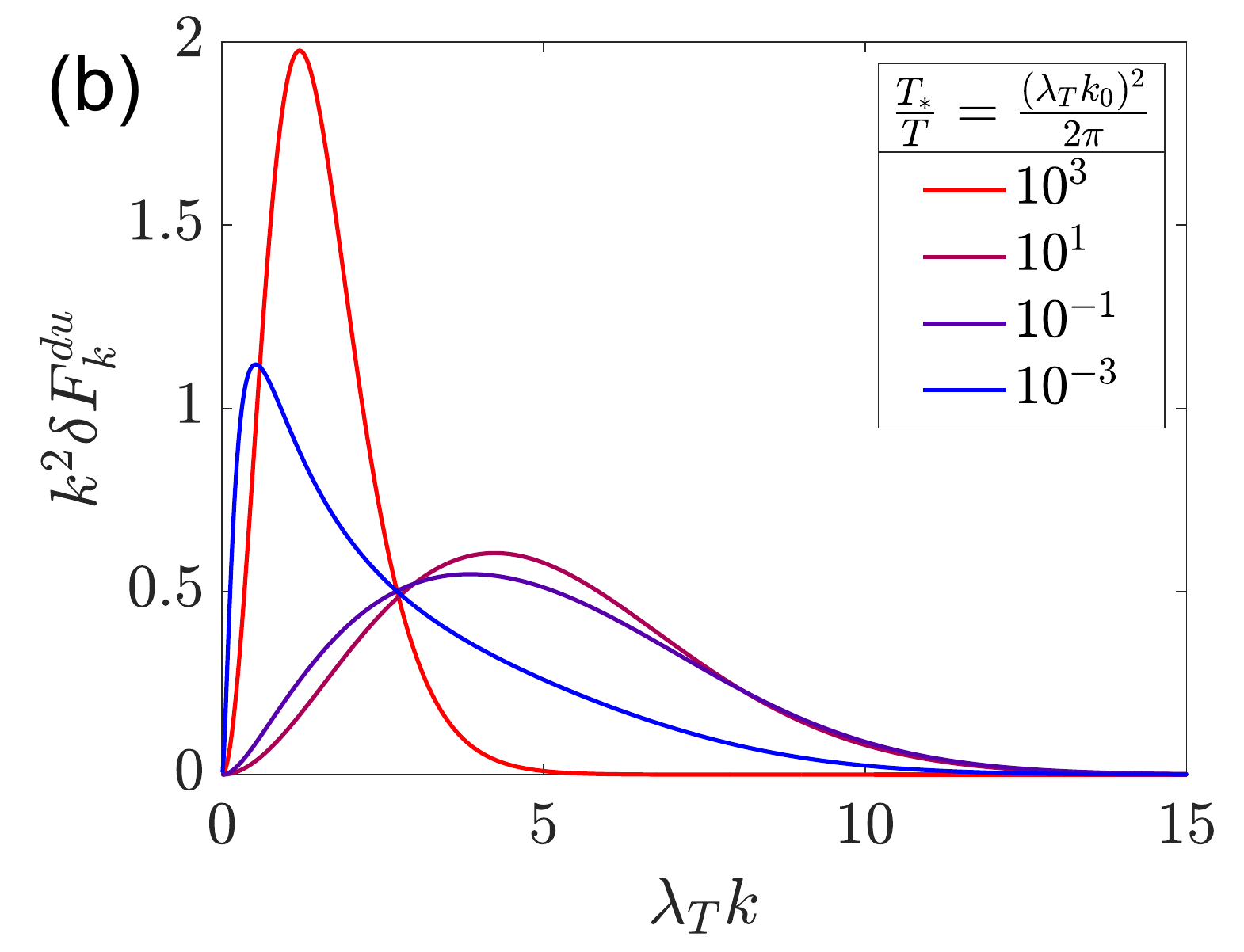}
\caption{\label{fig:lyap} (a) Blue solid line: Lyapunov exponent $\lambda_L$ as a function of temperature $T$. The $y$ axis is normalized to indicate the behavior $\lambda_L\propto T$ at $T\gg T_*$, while the $T\ll T_*$ behavior \eqref{eq:T5} is plotted in the red dashed line. Here we took a cutoff $\tilde{k}_{\mathrm{cut}}\lesssim 10$ in \eqref{eq:MF} and discretized to $n_{\mathrm{cut}}= 4000$ points of $\tilde{k}\le \tilde{k}_{\mathrm{cut}}$. We also computed the data when $\tilde{k}_{\mathrm{cut}}$ and $n_{\mathrm{cut}}$ is cut in half to estimate the error bar. The value of $\tilde{k}_{\mathrm{cut}}$ is optimized for each $T$, so that the error bar is barely visible. (b) The Lyapunov eigen-mode $\delta F^{du}_k$ as a function of $k$, at five temperatures shown in the legend. The amplitude of each mode is normalized so that $\int \dd k k^2 \delta F^{du}_k = 1$. A crucial observation is that $k^2 \delta F^{du}_k$ peaks at $k\sim k_0$ for $T\gg T_*$. }
\end{figure}

Since the Lyapunov exponent $\lambda_L$ characterizes local scrambling, we can assume the perturbation is homogeneous $F(x,\kk)=F_\kk(t)$ in space $\xx$. Assuming $F^{du}= F^{du}_0+\dF^{du}$ and expanding \eqref{eq:stud} to linear order in $\dF^{du}$, (\ref{eq:boltz}) becomes \begin{widetext} \begin{align}\label{eq:MF}
    \partial_t \dF^{du}_\kk &= 2\frac{g^2}{\hbar} \intq
    \Big\{ M^2_{\qq, \kk-\qq}\delta(\EE_\kk-\EE_\qq-\EE_{\kk-\qq})\mlr{F^{du}_0(\EE_{\kk-\qq})\dF^{du}_\qq  - F_0(\EE_\qq) \dF^{du}_\kk} \nonumber\\
    & \qquad\qquad\qquad +M^2_{\kk, \qq}\delta(\EE_\kk+\EE_\qq-\EE_{\kk+\qq})\mlr{F^{du}_0(\EE_{\qq}) \dF^{du}_{\kk+\qq}+F^{du}_0(\EE_{\kk+\qq})\dF^{du}_\qq  - \lr{F_0(\EE_\qq) - F_0(\EE_{\kk+\qq})} \dF^{du}_\kk}\Big\} \nonumber\\ &=\frac{8}{\hbar}\sqrt{na_s^3}\sqrt{TT_*}\int \frac{\dd^3\tqq}{\sqrt{2\pi}}
    \Bigg\{ \tilde{M}^2_{\tqq, \tkk-\tqq}\delta(\tE_\tkk-\tE_\tqq-\tE_{\tkk-\tqq})\mlr{\lr{\sinh\tE_{\tkk-\tqq}}^{-1}\dF^{du}_\tqq  - \coth\tE_\tqq \dF^{du}_\tkk} \nonumber\\
    & \qquad +\tilde{M}^2_{\tkk, \tqq}\delta(\tE_\tkk+\tE_\tqq-\tE_{\tkk+\tqq})\mlr{ \lr{\sinh\tE_{\tqq}}^{-1} \dF^{du}_{\tkk+\tqq}+\lr{\sinh\tE_{\tkk+\tqq}}^{-1}\dF^{du}_\tqq  - \lr{\coth\tE_\tqq - \coth\tE_{\tkk+\tqq}} \dF^{du}_\tkk}\Bigg\} \nonumber\\ &\equiv \sum_\tqq \mathcal{M}_{\tkk,\tqq} \dF^{du}_\tqq,
\end{align}
\end{widetext} 
where we have used the rescaled dimensionless parameters \begin{equation}\label{eq:tk}
    \tkk = \frac{\hbar \kk}{\sqrt{2mT}}, \quad \tE=\frac{\EE}{2T},\quad \tilde{M}=M/\sqrt{n}.
\end{equation}
The Lyapunov exponent is then the largest positive eigenvalue $\max \mathrm{eig}(\mathcal{M})$ of the matrix $\mathcal{M}$. Since $\tilde{M}$ and $\tilde{\EE}$ in \eqref{eq:MF} only depends on $T/T_*$, we have the general form \begin{equation}
    \lambda_L = \hbar^{-1}\sqrt{na_s^3}T_*\, f(T/T_*),
\end{equation} 
with $f(\cdot)$ being a universal function.
We further assume the mode corresponding to $\lambda_L$ is isotropic: $\dF_\kk(t) = \dF_k(t)$ with $k\equiv |\kk|$, so that \eqref{eq:MF} reduces to $\partial_t \dF_k = \sum_{k'}\bar{\mathcal{M}}_{k,k'}\dF_{k'}$, with details given in Appendix~\ref{sec:sym} on how to transform the integration measure. We take discrete values of $k$ up to a cutoff $k_{\rm cut}$ to generate the ${\cal M}$ matrix. The cutoff $k_{\rm cut}$ is much larger than $k_0$, such that the largest eigenvalues of ${\cal M}$ are approximately independent of $k_{\rm cut}$. Using the expression \eqref{eq:Mkk} for $M_{\kk,\qq}$, we then numerically solve for $\lambda_L=\max \mathrm{eig}(\bar{\mathcal{M}})$ as a function of $T$, as shown in Fig.~\ref{fig:lyap}(a).

At sufficiently low temperature $T\ll T_*$, the quasiparticles are typically phonon-like $\EE_\kk \approx \hbar ck$ for $k \ll k_0$. In this regime the collision matrix $M_{\kk_1,\kk_2}\approx 3 \sqrt{\frac{nk_1k_2k_3}{2^7k_0^3}}$ \cite{Matrix01}, so that the dependence of the matrix $\bar{\mathcal{M}}_{k,q}$ on $T/T_*$ can be extracted as a prefactor proportional to $T^5$. From numerics, we indeed get \begin{equation}\label{eq:T5}
    \lambda_L(T\ll T_*) \approx 761\hbar^{-1} \sqrt{na_s^3}T_*\lr{\frac{T}{T_*}}^5,
\end{equation}
as indicated by the red dashed line in Fig.~\ref{fig:lyap}(a). \eqref{eq:T5} agrees \emph{quantitatively} with the result on the unitary Fermi gas at low temperature that has a similar effective boson model \cite{Pengfei19}, validating our calculation. The $T^5$ scaling is parametrically smaller than the chaos bound \cite{Maldacena16a}. Because GBE share similar forms with traditional Boltzmann equations that govern damping of quasiparticles, one expect they have the same time scales. Indeed, \eqref{eq:T5} is of the same order as the Beliaev damping rate $\frac{1}{\tau(k)}\sim \frac{k^5}{\hbar^4 mn}$ \cite{beliaev58} evaluated at the typical phonon momentum $k\approx T/\hbar c$.

At relatively high temperature $T\gg T_*$, one can assume that all $k$ of interest are in the particle regime $k\sim \lambda_T^{-1}\gg k_0$ so that $M_{\kk_1,\kk_2}\approx \sqrt{n}$, and count the dimensions similarly. However, this naive dimension counting results in $\lambda_L\propto \sqrt{T}$, which disagrees with the numerical result \begin{equation}\label{eq:T1}
    \lambda_L(T\gg T_*) \approx 4\hbar^{-1}\sqrt{na_s^3}T.
\end{equation}
To resolve this issue, we plot the eigen-mode $\delta F^{du}_k$ that corresponds to the eigenvalue $\lambda_L$ in Fig.~\ref{fig:lyap}(b), where the low temperature case is also included. For the blue line $T/T_*=10^3$, We find that although the $k$ distribution $k^2 \delta F^{du}_k$ \footnote{Since $\delta F_{k}^{du}$ represents a density in $\bm k$ space and the system is isotropic, $k^2\delta F^{du}_k$ represents the density of $k$ with all angular directions being integrated.} sits largely in the $k\sim \lambda_T^{-1}$ regime, it peaks at $k\sim k_0$ instead. Thus the $k\lesssim k_0$ momentums also contribute nontrivially to $\lambda_T$, resulting in the failure of the naive dimension counting argument. Comparing \eqref{eq:T1} to the Landau damping rate $\frac{1}{\tau(k)}\sim \EE_k \frac{aT}{\hbar c}$ at this temperature region \cite{LandauD_74,damp09}, we find agreement $\lambda_L\sim \frac{1}{\tau(k)}$ only for the peak value $k\sim k_0$, instead of the typical one $k\sim \lambda_T^{-1}$. This shows an interesting phenomenon where information is mostly scrambled by the small fraction of low-energy quasiparticles.
The linear $T$ behavior in \eqref{eq:T1} mimics models with holographic duals \cite{Kitaev15,Maldacena16b}, although here the small prefactor $\sqrt{na_s^3}\ll 1$ means our theory is weakly interacting, and $\lambda_L$ is still parametrically smaller than the chaos bound \cite{Maldacena16a}.

\section{butterfly velocity}\label{sec:vB}

\begin{figure}[t]
\centering
\includegraphics[width=0.45\textwidth]{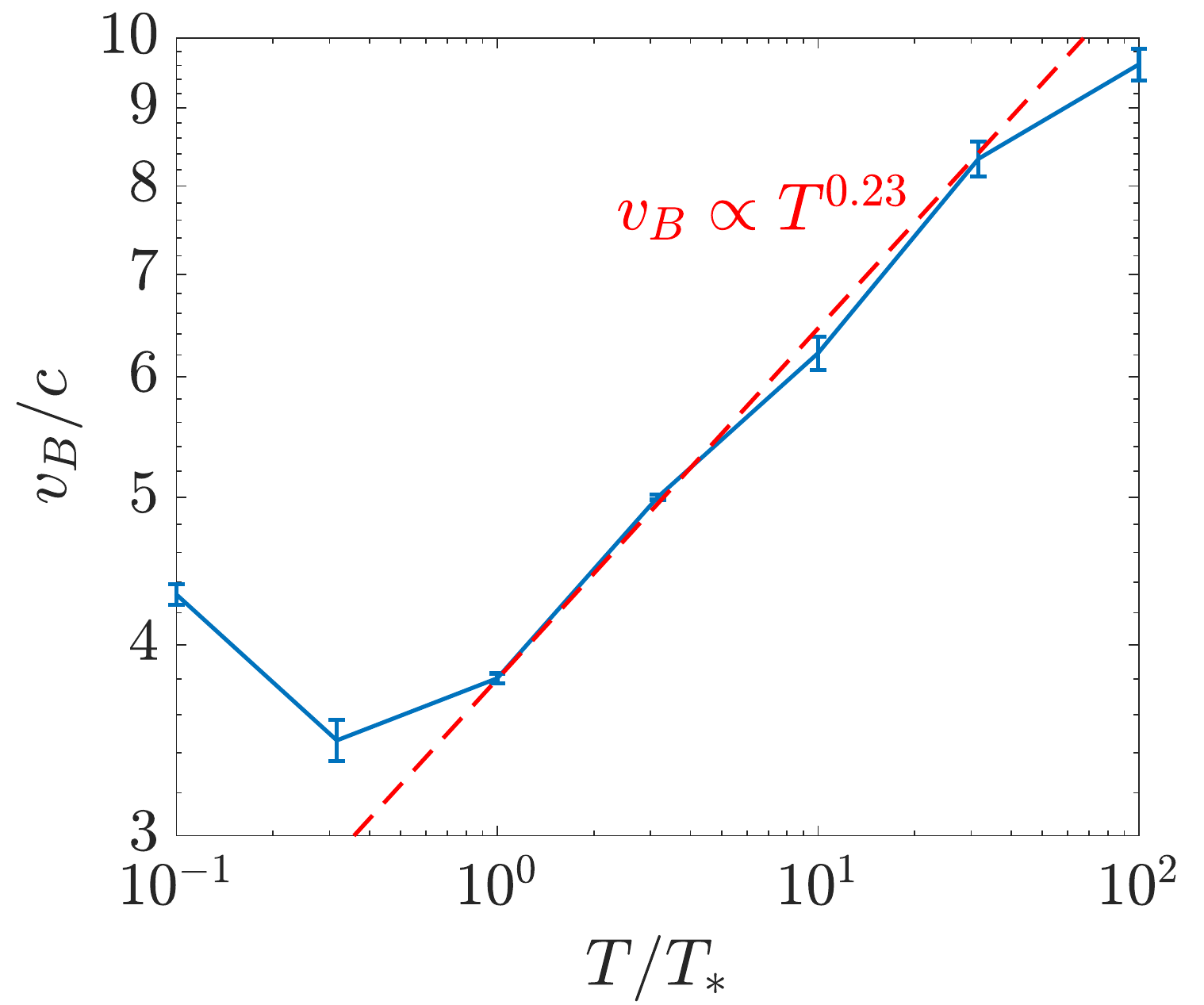}
\caption{\label{fig:v_B}  Butterfly velocity $v_B$ as a function of $T$ shown by blue solid line, while the red dashed line represents \eqref{eq:vT>}. { We assume rotational symmetry along $z$-axis that $\bm{\ell}$ points, and diagonalize \eqref{eq:M+lv} in the $(\tilde{k}_x=\tilde{k}\sin\theta_{\tilde{k}},\tilde{k}_z=\tilde{k}\cos\theta_{\tilde{k}})$ plane using \eqref{eq:int_meas} as the integration measure. }  We choose the region $0\le \tilde{k}_x \le \tilde{k}_{\mathrm{cut}}, -\tilde{k}_{\mathrm{cut}} \le \tilde{k}_z \le \tilde{k}_{\mathrm{cut}}$, with the cutoff $\tilde{k}_{\mathrm{cut}}\lesssim 4.5$ optimized for each $T$. This 2d region is discretized to $n_{\mathrm{cut}}=45000$ points, and we also computed the data when $\tilde{k}_{\mathrm{cut}}$ and $n_{\mathrm{cut}}$ are decreased by a factor of $4/5$, to obtain the error bar. }
\end{figure}

To further calculate the butterfly velocity $v_B$, we take the ansatz $\delta F^{du}(\xx)\propto \ee^{-\ii\bm{\ell}\cdot \xx}$ instead of the homogeneous one. Then the linearized GBE becomes
\begin{equation}\label{eq:M+lv}
\partial_t \delta F^{du}_\tkk = \sum_\tqq \lr{\mathcal{M}+\ii\bm{\ell}\cdot \vv}_{\tkk,\tqq} \dF^{du}_\tqq,
\end{equation}
with $\mathcal{M}$ the same as \eqref{eq:MF}, and the diagonal matrix \begin{equation}
    \vv_{\tkk,\tqq} = \vv_\kk \delta_{\tkk,\tqq},\quad \mathrm{where} \quad \vv_\kk=\nabla_\kk \EE_\kk/\hbar.
\end{equation}
The maximum eigenvalue
$\max\mathrm{eig}(M+\ii\bm{\ell}\cdot \vv)$ is then the Lyapunov exponent $\lambda_L(\ell)$ at wave vector $\ell$, and the general solution takes the form $\delta F^{du}\sim \int \dd\bm{\ell} \chi_{\bm{\ell}} \ee^{\lambda_L(\ell)t -\ii\bm{\ell}\cdot \xx}$. Suppose the initial perturbation varies slowly in space, so that we can expand at small $\ell$:
\begin{equation}
    \lambda_L(\ell) \approx \lambda_0 - \lambda_2 \ell^2 \pm \ii \lambda_1 \ell,
\end{equation}
where $\lambda_j$s are all non-negative. We then integrate over $\bm{\ell}$ by saddle-point approximation, { by finding $\bm{\ell}$ that satisfies
\begin{eqnarray}
\partial_\ell(\lambda_L(\ell)-\ii\bm{\ell}\cdot\bm{x})&=0,\\
\partial_\theta(\lambda_L(\ell)-\ii\bm{\ell}\cdot\bm{x})&=0,
\end{eqnarray}
where $\theta$ is the angle between ${\bm \ell}$ and ${\bm x}$. This gives}
\begin{equation}
    \delta F^{du}\sim \exp\mlr{ \lambda_0t -\frac{(|\xx|-\lambda_1 t)^2}{4\lambda_2 t}},
\end{equation}
which decays exponentially for $|\xx|>v_B t$, where
\begin{equation}
    v_B = \lambda_1 + 2\sqrt{\lambda_0\lambda_2} \equiv c\tilde{f}(T/T_*),
\end{equation}
for some universal function $\tilde{f}(\cdot)$.

\begin{table}[b]
\caption{\label{tab}
Scaling laws of physical quantities describing chaos and energy diffusion, in the two temperature regimes.
}
\begin{ruledtabular}
\begin{tabular}{ccc}
physical quantity& $T\ll T_*$ & $T\gg T_*$ \\
\colrule
$\lambda_L$ & $\hbar^{-1}\sqrt{na_s^3}T_*\lr{\frac{T}{T_*}}^5$ & $\hbar^{-1}\sqrt{na_s^3} T$ \\
$v_B$ & $c$ & $c\lr{\frac{T}{T_*}}^{0.23}$ \\
$D_L=v_B^2/\lambda_L$ & $\frac{\hbar}{m\sqrt{na^3}}\lr{\frac{T_*}{T}}^5$ & $\frac{\hbar}{m\sqrt{na^3}}\lr{\frac{T_*}{T}}^{0.54}$ \\
$D_E=\kappa/c_v$ & $\frac{\hbar}{m\sqrt{na^3}}\frac{T_*}{T}$ & $\frac{\hbar}{m\sqrt{na^3}}\sqrt{\frac{T}{T_*}}$ \\
\end{tabular}
\end{ruledtabular}
\end{table}

We numerically calculate $\lambda_j$s by diagonalizing $M+\ii\bm{\ell}\cdot \vv$, and get $v_B/c$ as a function of $T/T_*$ in Fig.~\ref{fig:v_B}. Following the dimension-counting arguments in the previous section, at very low temperature $T\ll T_*$, $v_B$ is several times of the sound velocity $c$, the only velocity scale present in the system. Fig.~\ref{fig:v_B} suggests \begin{equation}
    v_B(T\ll T_*) \approx 4 c.
\end{equation}
However, the numerical factor $4$ may be modified at temperature lower than $0.1T_*$, where our numerical algorithm yields fluctuating results and is thus not reliable. The more interesting region is at relatively high temperature $T\gg T_*$, where simple dimension counting fails. In Fig.~\ref{fig:v_B}, we observe a power law dependence \begin{equation}\label{eq:vT>}
    v_B(T\gg T_*) \approx 3.8 c \lr{\frac{T}{T_*}}^{0.23},
\end{equation}
which is parametrically smaller than the typical velocity $\sqrt{2T/m}$ of quasiparticles. This should be related to the anomalous clustering of the distribution $\delta F^{du}_k$ at small $k\sim k_0$ in Fig.~\ref{fig:lyap}(b), and demands further understanding. Surprisingly, the exponent $0.23$ matches the one for the butterfly velocity in a classical spin chain \cite{classical_chain_vb}, which suggests that drastically different microscopic models, may share universal behaviors regarding information scrambling dynamics.

Using the values of $\lambda_L$ and $v_B$, we calculate the quantity \begin{equation}
    D_L=v_B^2/\lambda_L \sim \frac{\hbar}{m\sqrt{na^3}} \times \left\{\begin{array}{lc}
        \lr{\frac{T_*}{T}}^5, & T\ll T_* \\
        \lr{\frac{T_*}{T}}^{0.54}, & T\gg T_*
    \end{array}\right. .
\end{equation}
In certain strongly interacting models \cite{E_diff,YIngfei17a,Sachdev17b}, such a chaos diffusion constant is found to agree with charge \cite{Blake_prl,Blake_prd} and energy \cite{E_diff} diffusion constants. However, the model we study here is weakly interacting, and one does not expect $D_L$ is related to the energy diffusion constant $D_E=\kappa/c_v$ \cite{Lucas17,otoc_ele_phon}. 
Indeed, using the formulas for heat conductivity $\kappa$ \cite{kirk83,kirk85} and heat capacity $c_v$ \cite{BEC_book}, we get $D_E$ for the dilute Bose gas: \begin{equation}
    D_E \sim \frac{\hbar}{m\sqrt{na^3}} \times \left\{\begin{array}{lc}
        \frac{T_*}{T}, & T\ll T_* \\
        \sqrt{\frac{T}{T_*}}, & T\gg T_*
    \end{array}\right. ,
\end{equation}
which is not equal to $D_L$. Remarkably, both possibilities $D_E\ll D_L$ and $D_E\gg D_L$ arise, at very low and relatively high temperatures respectively.

\section{conclusion}\label{sec:conclude}
In this paper, we have calculated the quantum Lyapounov exponent $\lambda_L$ and butterfly velocity $v_B$ of the dilute Bose gas in the BEC phase, with results summarized in Table \ref{tab}. We find $\lambda_L\propto T^5$ at very low temperature $T\ll T_*$ and $\lambda_L\propto T$ at relatively high temperature $T_*\ll T \ll T_{\rm BEC}$. Meanwhile, we find $v_B$ is at the order of the sound speed $c$ at very low temperature, and follows a $T^{0.23}$ power law at relatively high temperature. We have compared $\lambda_L$ with the quasiparticle damping rate, and the chaos diffusion constant $D_L=v_B^2/\lambda_L$ with the energy diffusion constant $D_E$. { The weakly interacting nature of the model is manifested by the asymptotic smallness of $\lambda_L$ compared to the chaos bound, and the mismatch between $D_L$ and $D_E$.} Our GBE method is proved to be efficient for calculating OTOC, since only 2-point functions are involved.  Experimental tests of our predictions would require either approaches to measure OTOC directly \cite{measure_otoc16,measure_otoc19}, or phenomenological connections between information scrambling and time-ordered physics. On the other hand, we expect our results can be generalized to higher temperature $T\sim T_{\rm BEC}$, where fluctuation of the condensate and vortices become important \cite{BEC_T_book}.

\emph{Acknowledgements.}--- We thank Pengfei Zhang and Andrew Lucas for useful discussions.  Y. C is supported by Beijing Natural Science Foundation (Z180013), and NSFC under Grant No. 12174358 and No. 11734010.

\begin{appendix}
\begin{widetext}

\section{integration measure with spherical or axial symmetry}\label{sec:sym}
We work with the dimensionless momentum $\tkk$ defined in \eqref{eq:tk}. First, assume $\delta F^{du}_\tkk = \delta F^{du}_{\tk}$ has spherical symmetry $\mathrm{SO}(3)$. Using $(\tq,\theta,\varphi)$ as the spherical coordinate of $\tqq$ with the polar axis pointing along $\tkk$, the integration measure in \eqref{eq:MF} becomes \begin{equation}\label{eq:deltatheta}
    \mathcal{I}\equiv \int \frac{\dd^3\tqq}{\sqrt{2\pi}}
    \delta(\tE_\tkk\mp \tE_\tqq-\tE_{\tkk\mp \tqq}) = \sqrt{2\pi} \int \tq^2\dd \tq \sin\theta \dd\theta \frac{\delta(\theta-\theta^{\mathrm{os}})}{\ilr{\partial_\theta \tE_3}},
\end{equation}
where $\varphi$ has been integrated over, $\tE_3\equiv\tE_{\tkk\mp \tqq}$ is the energy of the third quasiparticle, and $\theta^{\mathrm{os}}$ is the polar angle such that the corresponding $\tqq^{\mathrm{os}}$ with length $|\tqq^{\mathrm{os}}|=\tilde{q}$ makes the three quasiparticles on-shell: $\tE_\tkk\mp \tE_{\tqq^{\mathrm{os}}}-\tE_{\tkk\mp \tqq^{\mathrm{os}}}=0$. To calculate the denominator in \eqref{eq:deltatheta}, we use \begin{equation}\label{eq:dtheta}
    \partial_\theta = \lr{\partial_\theta \tq_3^2} \ddd{}{\tq_3^2} = \pm\frac{\tk\tq}{\tq_3}\sin \theta \ddd{}{\tq_3},
\end{equation}
because the momentum for the third quasiparticle is \begin{equation}
    \tq_3^2 = \tk^2+\tq^2 \mp 2\tk\tq \cos\theta.
\end{equation}
Then we integrate over $\theta$ in \eqref{eq:deltatheta} to get the spherical symmetric integration measure \begin{equation}
    \mathcal{I} = \sqrt{2\pi} \int^\infty_0 \dd \tq\frac{\tq_3\tq}{\tk \ilr{\ddd{\tE_3}{\tq_3}}},
\end{equation}
where $\tq_3(\tk,\tq)$ is the on-shell momentum such that \begin{equation}\label{eq:q3os}
    \tE_3(\tq_3)=\tE_\tk \mp \tE_\tq.
\end{equation}

More generally, assume $\delta F^{du}_\tkk = \delta F^{du}_{\tk,\theta_\tk}$ is not spherical symmetric, but still has the axial rotation symmetry $\mathrm{SO}(2)$ around the polar axis. In this case, assuming $\tkk$ correspond to $\varphi_\tk=0$, \eqref{eq:deltatheta} becomes \begin{equation}
    \mathcal{I} = \int \frac{\tq^2\dd\tq \sin\theta \dd \theta}{\sqrt{2\pi}} \sum_j\frac{\delta(\varphi-\varphi^{\mathrm{os}}_j)}{\ilr{\partial_\varphi \tE_3}},
\end{equation}
where there are either two or zero on-shell solutions for $\varphi^{\mathrm{os}}_j$. The third momentum is now \begin{equation}
    \tq_3^2 = \tk^2+\tq^2 \mp 2\tk\tq \lr{\cos\theta_\tk\cos\theta+\sin\theta_\tk\sin\theta \cos\varphi},
\end{equation}
so that \begin{equation}
    \ilr{\partial_\varphi \tq_3^2 }= \ilr{2\tk\tq\sin\theta_\tk\sin\theta \sin\varphi} = 2\tk\tq \mlr{\sin^2\theta_\tk\sin^2\theta - \lr{\frac{\tq_3^2-\tk^2-\tq^2}{2\tk\tq}\pm \cos\theta_\tk\cos\theta}^2}^{1/2}.
\end{equation}
Finally, we follow the similar strategy in \eqref{eq:dtheta} to get \begin{equation}\label{eq:int_meas}
    \mathcal{I} = 2\int \frac{\dd\tq\dd\theta}{\sqrt{2\pi}} \frac{\tq\tq_3\sin\theta}{\tk \ilr{\ddd{\tE_3}{\tq_3}}} \mlr{\sin^2\theta_\tk\sin^2\theta - \lr{\frac{\tq_3^2-\tk^2-\tq^2}{2\tk\tq}\pm \cos\theta_\tk\cos\theta}^2}^{-1/2},
\end{equation}
where $\tq_3$ is the on-shell solution for \eqref{eq:q3os}.

\end{widetext}

\end{appendix}

\bibliography{gBE_OTOC}

\providecommand{\noopsort}[1]{}\providecommand{\singleletter}[1]{#1}%
\begin{thebibliography}{69}%
\makeatletter
\providecommand \@ifxundefined [1]{%
 \@ifx{#1\undefined}
}%
\providecommand \@ifnum [1]{%
 \ifnum #1\expandafter \@firstoftwo
 \else \expandafter \@secondoftwo
 \fi
}%
\providecommand \@ifx [1]{%
 \ifx #1\expandafter \@firstoftwo
 \else \expandafter \@secondoftwo
 \fi
}%
\providecommand \natexlab [1]{#1}%
\providecommand \enquote  [1]{``#1''}%
\providecommand \bibnamefont  [1]{#1}%
\providecommand \bibfnamefont [1]{#1}%
\providecommand \citenamefont [1]{#1}%
\providecommand \href@noop [0]{\@secondoftwo}%
\providecommand \href [0]{\begingroup \@sanitize@url \@href}%
\providecommand \@href[1]{\@@startlink{#1}\@@href}%
\providecommand \@@href[1]{\endgroup#1\@@endlink}%
\providecommand \@sanitize@url [0]{\catcode `\\12\catcode `\$12\catcode
  `\&12\catcode `\#12\catcode `\^12\catcode `\_12\catcode `\%12\relax}%
\providecommand \@@startlink[1]{}%
\providecommand \@@endlink[0]{}%
\providecommand \url  [0]{\begingroup\@sanitize@url \@url }%
\providecommand \@url [1]{\endgroup\@href {#1}{\urlprefix }}%
\providecommand \urlprefix  [0]{URL }%
\providecommand \Eprint [0]{\href }%
\providecommand \doibase [0]{http://dx.doi.org/}%
\providecommand \selectlanguage [0]{\@gobble}%
\providecommand \bibinfo  [0]{\@secondoftwo}%
\providecommand \bibfield  [0]{\@secondoftwo}%
\providecommand \translation [1]{[#1]}%
\providecommand \BibitemOpen [0]{}%
\providecommand \bibitemStop [0]{}%
\providecommand \bibitemNoStop [0]{.\EOS\space}%
\providecommand \EOS [0]{\spacefactor3000\relax}%
\providecommand \BibitemShut  [1]{\csname bibitem#1\endcsname}%
\let\auto@bib@innerbib\@empty
%</preamble>
\bibitem [{\citenamefont {Larkin}\ and\ \citenamefont
  {Ovchinnikov}(1969)}]{LO69}%
  \BibitemOpen
  \bibfield  {author} {\bibinfo {author} {\bibfnamefont {A.}~\bibnamefont
  {Larkin}}\ and\ \bibinfo {author} {\bibfnamefont {Y.~N.}\ \bibnamefont
  {Ovchinnikov}},\ }\bibfield  {title} {\enquote {\bibinfo {title}
  {Quasiclassical method in the theory of superconductivity},}\ }\href@noop {}
  {\bibfield  {journal} {\bibinfo  {journal} {JETP}\ }\textbf {\bibinfo
  {volume} {28}},\ \bibinfo {pages} {960} (\bibinfo {year} {1969})}\BibitemShut
  {NoStop}%
\bibitem [{\citenamefont {Kitaev}(2014)}]{Kitaev14}%
  \BibitemOpen
  \bibfield  {author} {\bibinfo {author} {\bibfnamefont {A.}~\bibnamefont
  {Kitaev}},\ }\bibfield  {title} {\enquote {\bibinfo {title} {talk given at
  fundamental physics prize symposium},}\ }\href
  {http://online.kitp.ucsb.edu/online/joint98/kitaev/} {\bibfield  {journal}
  {\bibinfo  {journal} {http://online.kitp.ucsb.edu/online/joint98/kitaev/}\ }
  (\bibinfo {year} {2014})}\BibitemShut {NoStop}%
\bibitem [{\citenamefont {Shenker}\ and\ \citenamefont
  {Stanford}(2014{\natexlab{a}})}]{Shenker14}%
  \BibitemOpen
  \bibfield  {author} {\bibinfo {author} {\bibfnamefont {Stephen~H.}\
  \bibnamefont {Shenker}}\ and\ \bibinfo {author} {\bibfnamefont
  {D.}~\bibnamefont {Stanford}},\ }\bibfield  {title} {\enquote {\bibinfo
  {title} {Black holes and the butterfly effect},}\ }\href
  {https://doi.org/10.1007/JHEP03(2014)067} {\bibfield  {journal} {\bibinfo
  {journal} {Journal of High Energy Physics}\ }\textbf {\bibinfo {volume}
  {03}},\ \bibinfo {pages} {067} (\bibinfo {year}
  {2014}{\natexlab{a}})}\BibitemShut {NoStop}%
\bibitem [{\citenamefont {Rozenbaum}\ \emph {et~al.}(2017)\citenamefont
  {Rozenbaum}, \citenamefont {Ganeshan},\ and\ \citenamefont
  {Galitski}}]{Galitski17}%
  \BibitemOpen
  \bibfield  {author} {\bibinfo {author} {\bibfnamefont {Efim~B.}\ \bibnamefont
  {Rozenbaum}}, \bibinfo {author} {\bibfnamefont {Sriram}\ \bibnamefont
  {Ganeshan}}, \ and\ \bibinfo {author} {\bibfnamefont {Victor}\ \bibnamefont
  {Galitski}},\ }\bibfield  {title} {\enquote {\bibinfo {title} {Lyapunov
  exponent and out-of-time-ordered correlator's growth rate in a chaotic
  system},}\ }\href {\doibase 10.1103/PhysRevLett.118.086801} {\bibfield
  {journal} {\bibinfo  {journal} {Phys. Rev. Lett.}\ }\textbf {\bibinfo
  {volume} {118}},\ \bibinfo {pages} {086801} (\bibinfo {year}
  {2017})}\BibitemShut {NoStop}%
\bibitem [{\citenamefont {Xu}\ \emph {et~al.}(2020)\citenamefont {Xu},
  \citenamefont {Scaffidi},\ and\ \citenamefont {Cao}}]{Xu20}%
  \BibitemOpen
  \bibfield  {author} {\bibinfo {author} {\bibfnamefont {Tianrui}\ \bibnamefont
  {Xu}}, \bibinfo {author} {\bibfnamefont {Thomas}\ \bibnamefont {Scaffidi}}, \
  and\ \bibinfo {author} {\bibfnamefont {Xiangyu}\ \bibnamefont {Cao}},\
  }\bibfield  {title} {\enquote {\bibinfo {title} {Does scrambling equal
  chaos?}}\ }\href {\doibase 10.1103/PhysRevLett.124.140602} {\bibfield
  {journal} {\bibinfo  {journal} {Phys. Rev. Lett.}\ }\textbf {\bibinfo
  {volume} {124}},\ \bibinfo {pages} {140602} (\bibinfo {year}
  {2020})}\BibitemShut {NoStop}%
\bibitem [{\citenamefont {Yin}\ and\ \citenamefont {Lucas}(2021)}]{Yin_top}%
  \BibitemOpen
  \bibfield  {author} {\bibinfo {author} {\bibfnamefont {Chao}\ \bibnamefont
  {Yin}}\ and\ \bibinfo {author} {\bibfnamefont {Andrew}\ \bibnamefont
  {Lucas}},\ }\bibfield  {title} {\enquote {\bibinfo {title} {Quantum operator
  growth bounds for kicked tops and semiclassical spin chains},}\ }\href
  {\doibase 10.1103/PhysRevA.103.042414} {\bibfield  {journal} {\bibinfo
  {journal} {Phys. Rev. A}\ }\textbf {\bibinfo {volume} {103}},\ \bibinfo
  {pages} {042414} (\bibinfo {year} {2021})}\BibitemShut {NoStop}%
\bibitem [{\citenamefont {Maldacena}\ \emph
  {et~al.}(2016{\natexlab{a}})\citenamefont {Maldacena}, \citenamefont
  {Shenker},\ and\ \citenamefont {Stanford}}]{Maldacena16a}%
  \BibitemOpen
  \bibfield  {author} {\bibinfo {author} {\bibfnamefont {Juan}\ \bibnamefont
  {Maldacena}}, \bibinfo {author} {\bibfnamefont {Stephen~H.}\ \bibnamefont
  {Shenker}}, \ and\ \bibinfo {author} {\bibfnamefont {Douglas}\ \bibnamefont
  {Stanford}},\ }\bibfield  {title} {\enquote {\bibinfo {title} {{A bound on
  chaos}},}\ }\href {\doibase 10.1007/JHEP08(2016)106} {\bibfield  {journal}
  {\bibinfo  {journal} {JHEP}\ }\textbf {\bibinfo {volume} {08}},\ \bibinfo
  {pages} {106} (\bibinfo {year} {2016}{\natexlab{a}})},\ \Eprint
  {http://arxiv.org/abs/1503.01409} {arXiv:1503.01409 [hep-th]} \BibitemShut
  {NoStop}%
\bibitem [{\citenamefont {Shenker}\ and\ \citenamefont
  {Stanford}(2014{\natexlab{b}})}]{Shenker14b}%
  \BibitemOpen
  \bibfield  {author} {\bibinfo {author} {\bibfnamefont {Stephen~H.}\
  \bibnamefont {Shenker}}\ and\ \bibinfo {author} {\bibfnamefont {Douglas}\
  \bibnamefont {Stanford}},\ }\bibfield  {title} {\enquote {\bibinfo {title}
  {{Multiple Shocks}},}\ }\href {\doibase 10.1007/JHEP12(2014)046} {\bibfield
  {journal} {\bibinfo  {journal} {JHEP}\ }\textbf {\bibinfo {volume} {12}},\
  \bibinfo {pages} {046} (\bibinfo {year} {2014}{\natexlab{b}})},\ \Eprint
  {http://arxiv.org/abs/1312.3296} {arXiv:1312.3296 [hep-th]} \BibitemShut
  {NoStop}%
\bibitem [{\citenamefont {Shenker}\ and\ \citenamefont
  {Stanford}(2015)}]{Shenker15}%
  \BibitemOpen
  \bibfield  {author} {\bibinfo {author} {\bibfnamefont {Stephen~H.}\
  \bibnamefont {Shenker}}\ and\ \bibinfo {author} {\bibfnamefont {Douglas}\
  \bibnamefont {Stanford}},\ }\bibfield  {title} {\enquote {\bibinfo {title}
  {{Stringy effects in scrambling}},}\ }\href {\doibase
  10.1007/JHEP05(2015)132} {\bibfield  {journal} {\bibinfo  {journal} {JHEP}\
  }\textbf {\bibinfo {volume} {05}},\ \bibinfo {pages} {132} (\bibinfo {year}
  {2015})},\ \Eprint {http://arxiv.org/abs/1412.6087} {arXiv:1412.6087
  [hep-th]} \BibitemShut {NoStop}%
\bibitem [{\citenamefont {Roberts}\ \emph {et~al.}(2015)\citenamefont
  {Roberts}, \citenamefont {Stanford},\ and\ \citenamefont
  {Susskind}}]{Susskind15}%
  \BibitemOpen
  \bibfield  {author} {\bibinfo {author} {\bibfnamefont {Daniel~A.}\
  \bibnamefont {Roberts}}, \bibinfo {author} {\bibfnamefont {Douglas}\
  \bibnamefont {Stanford}}, \ and\ \bibinfo {author} {\bibfnamefont {Leonard}\
  \bibnamefont {Susskind}},\ }\bibfield  {title} {\enquote {\bibinfo {title}
  {{Localized shocks}},}\ }\href {\doibase 10.1007/JHEP03(2015)051} {\bibfield
  {journal} {\bibinfo  {journal} {JHEP}\ }\textbf {\bibinfo {volume} {03}},\
  \bibinfo {pages} {051} (\bibinfo {year} {2015})},\ \Eprint
  {http://arxiv.org/abs/1409.8180} {arXiv:1409.8180 [hep-th]} \BibitemShut
  {NoStop}%
\bibitem [{\citenamefont {Kitaev}(2015)}]{Kitaev15}%
  \BibitemOpen
  \bibfield  {author} {\bibinfo {author} {\bibfnamefont {A.}~\bibnamefont
  {Kitaev}},\ }\bibfield  {title} {\enquote {\bibinfo {title} {talk given at
  kitp program: Entanglement in strongly-correlated quantum matter},}\ }\href
  {http://online.kitp.ucsb.edu/online/entangled15/kitaev/} {\bibfield
  {journal} {\bibinfo  {journal}
  {http://online.kitp.ucsb.edu/online/entangled15/kitaev/}\ } (\bibinfo {year}
  {2015})}\BibitemShut {NoStop}%
\bibitem [{\citenamefont {Maldacena}\ and\ \citenamefont
  {Stanford}(2016)}]{Maldacena16b}%
  \BibitemOpen
  \bibfield  {author} {\bibinfo {author} {\bibfnamefont {Juan}\ \bibnamefont
  {Maldacena}}\ and\ \bibinfo {author} {\bibfnamefont {Douglas}\ \bibnamefont
  {Stanford}},\ }\bibfield  {title} {\enquote {\bibinfo {title} {Remarks on the
  sachdev-ye-kitaev model},}\ }\href {\doibase 10.1103/PhysRevD.94.106002}
  {\bibfield  {journal} {\bibinfo  {journal} {Phys. Rev. D}\ }\textbf {\bibinfo
  {volume} {94}},\ \bibinfo {pages} {106002} (\bibinfo {year}
  {2016})}\BibitemShut {NoStop}%
\bibitem [{\citenamefont {Maldacena}\ \emph
  {et~al.}(2016{\natexlab{b}})\citenamefont {Maldacena}, \citenamefont
  {Stanford},\ and\ \citenamefont {Yang}}]{Maldacena16c}%
  \BibitemOpen
  \bibfield  {author} {\bibinfo {author} {\bibfnamefont {Juan}\ \bibnamefont
  {Maldacena}}, \bibinfo {author} {\bibfnamefont {Douglas}\ \bibnamefont
  {Stanford}}, \ and\ \bibinfo {author} {\bibfnamefont {Zhenbin}\ \bibnamefont
  {Yang}},\ }\bibfield  {title} {\enquote {\bibinfo {title} {{Conformal
  symmetry and its breaking in two-dimensional nearly anti-de Sitter space}},}\
  }\href {http://doi.org/10.1093/ptep/ptw124} {\bibfield  {journal} {\bibinfo
  {journal} {Progress of Theoretical and Experimental Physics}\ }\textbf
  {\bibinfo {volume} {2016}} (\bibinfo {year}
  {2016}{\natexlab{b}})}\BibitemShut {NoStop}%
\bibitem [{\citenamefont {Jackiw}(1985)}]{Jackiw85}%
  \BibitemOpen
  \bibfield  {author} {\bibinfo {author} {\bibfnamefont {R.}~\bibnamefont
  {Jackiw}},\ }\bibfield  {title} {\enquote {\bibinfo {title} {Lower
  dimensional gravity},}\ }\href
  {http://www.sciencedirect.com/science/article/pii/0550321385904481}
  {\bibfield  {journal} {\bibinfo  {journal} {Nuclear Physics B}\ }\textbf
  {\bibinfo {volume} {252}},\ \bibinfo {pages} {343--356} (\bibinfo {year}
  {1985})}\BibitemShut {NoStop}%
\bibitem [{\citenamefont {Teitelboim}(1983)}]{Teitelboim83}%
  \BibitemOpen
  \bibfield  {author} {\bibinfo {author} {\bibfnamefont {Claudio}\ \bibnamefont
  {Teitelboim}},\ }\bibfield  {title} {\enquote {\bibinfo {title} {Gravitation
  and hamiltonian structure in two spacetime dimensions},}\ }\href
  {http://www.sciencedirect.com/science/article/pii/0370269383900126}
  {\bibfield  {journal} {\bibinfo  {journal} {Physics Letters B}\ }\textbf
  {\bibinfo {volume} {126}},\ \bibinfo {pages} {41--45} (\bibinfo {year}
  {1983})}\BibitemShut {NoStop}%
\bibitem [{\citenamefont {Lashkari}\ \emph {et~al.}(2013)\citenamefont
  {Lashkari}, \citenamefont {Stanford}, \citenamefont {Hastings}, \citenamefont
  {Osborne},\ and\ \citenamefont {Hayden}}]{fast_scram13}%
  \BibitemOpen
  \bibfield  {author} {\bibinfo {author} {\bibfnamefont {Nima}\ \bibnamefont
  {Lashkari}}, \bibinfo {author} {\bibfnamefont {Douglas}\ \bibnamefont
  {Stanford}}, \bibinfo {author} {\bibfnamefont {Matthew}\ \bibnamefont
  {Hastings}}, \bibinfo {author} {\bibfnamefont {Tobias}\ \bibnamefont
  {Osborne}}, \ and\ \bibinfo {author} {\bibfnamefont {Patrick}\ \bibnamefont
  {Hayden}},\ }\bibfield  {title} {\enquote {\bibinfo {title} {{Towards the
  Fast Scrambling Conjecture}},}\ }\href {\doibase 10.1007/JHEP04(2013)022}
  {\bibfield  {journal} {\bibinfo  {journal} {JHEP}\ }\textbf {\bibinfo
  {volume} {04}},\ \bibinfo {pages} {022} (\bibinfo {year} {2013})},\ \Eprint
  {http://arxiv.org/abs/1111.6580} {arXiv:1111.6580 [hep-th]} \BibitemShut
  {NoStop}%
\bibitem [{\citenamefont {Bentsen}\ \emph {et~al.}(2019)\citenamefont
  {Bentsen}, \citenamefont {Gu},\ and\ \citenamefont {Lucas}}]{fast_scram19}%
  \BibitemOpen
  \bibfield  {author} {\bibinfo {author} {\bibfnamefont {Gregory}\ \bibnamefont
  {Bentsen}}, \bibinfo {author} {\bibfnamefont {Yingfei}\ \bibnamefont {Gu}}, \
  and\ \bibinfo {author} {\bibfnamefont {Andrew}\ \bibnamefont {Lucas}},\
  }\bibfield  {title} {\enquote {\bibinfo {title} {Fast scrambling on sparse
  graphs},}\ }\href {\doibase 10.1073/pnas.1811033116} {\bibfield  {journal}
  {\bibinfo  {journal} {Proceedings of the National Academy of Sciences}\
  }\textbf {\bibinfo {volume} {116}},\ \bibinfo {pages} {6689–6694} (\bibinfo
  {year} {2019})}\BibitemShut {NoStop}%
\bibitem [{\citenamefont {Yin}\ and\ \citenamefont
  {Lucas}(2020)}]{Yin_all2all}%
  \BibitemOpen
  \bibfield  {author} {\bibinfo {author} {\bibfnamefont {Chao}\ \bibnamefont
  {Yin}}\ and\ \bibinfo {author} {\bibfnamefont {Andrew}\ \bibnamefont
  {Lucas}},\ }\bibfield  {title} {\enquote {\bibinfo {title} {Bound on quantum
  scrambling with all-to-all interactions},}\ }\href {\doibase
  10.1103/PhysRevA.102.022402} {\bibfield  {journal} {\bibinfo  {journal}
  {Phys. Rev. A}\ }\textbf {\bibinfo {volume} {102}},\ \bibinfo {pages}
  {022402} (\bibinfo {year} {2020})}\BibitemShut {NoStop}%
\bibitem [{\citenamefont {Hosur}\ \emph {et~al.}(2016)\citenamefont {Hosur},
  \citenamefont {Qi}, \citenamefont {Roberts},\ and\ \citenamefont
  {Yoshida}}]{Yoshida16}%
  \BibitemOpen
  \bibfield  {author} {\bibinfo {author} {\bibfnamefont {Pavan}\ \bibnamefont
  {Hosur}}, \bibinfo {author} {\bibfnamefont {Xiao-Liang}\ \bibnamefont {Qi}},
  \bibinfo {author} {\bibfnamefont {Daniel~A.}\ \bibnamefont {Roberts}}, \ and\
  \bibinfo {author} {\bibfnamefont {Beni}\ \bibnamefont {Yoshida}},\ }\bibfield
   {title} {\enquote {\bibinfo {title} {Chaos in quantum channels},}\ }\href
  {\doibase 10.1103/PhysRevLett.124.140602} {\bibfield  {journal} {\bibinfo
  {journal} {Journal of High Energy Physics}\ }\textbf {\bibinfo {volume}
  {04}},\ \bibinfo {pages} {02} (\bibinfo {year} {2016})}\BibitemShut {NoStop}%
\bibitem [{\citenamefont {Roberts}\ and\ \citenamefont
  {Swingle}(2016)}]{Swingle16}%
  \BibitemOpen
  \bibfield  {author} {\bibinfo {author} {\bibfnamefont {Daniel~A.}\
  \bibnamefont {Roberts}}\ and\ \bibinfo {author} {\bibfnamefont {Brian}\
  \bibnamefont {Swingle}},\ }\bibfield  {title} {\enquote {\bibinfo {title}
  {Lieb-robinson bound and the butterfly effect in quantum field theories},}\
  }\href {\doibase 10.1103/PhysRevLett.117.091602} {\bibfield  {journal}
  {\bibinfo  {journal} {Phys. Rev. Lett.}\ }\textbf {\bibinfo {volume} {117}},\
  \bibinfo {pages} {091602} (\bibinfo {year} {2016})}\BibitemShut {NoStop}%
\bibitem [{\citenamefont {Blake}(2016{\natexlab{a}})}]{Blake_prl}%
  \BibitemOpen
  \bibfield  {author} {\bibinfo {author} {\bibfnamefont {Mike}\ \bibnamefont
  {Blake}},\ }\bibfield  {title} {\enquote {\bibinfo {title} {Universal charge
  diffusion and the butterfly effect in holographic theories},}\ }\href
  {\doibase 10.1103/PhysRevLett.117.091601} {\bibfield  {journal} {\bibinfo
  {journal} {Phys. Rev. Lett.}\ }\textbf {\bibinfo {volume} {117}},\ \bibinfo
  {pages} {091601} (\bibinfo {year} {2016}{\natexlab{a}})}\BibitemShut
  {NoStop}%
\bibitem [{\citenamefont {Lucas}\ and\ \citenamefont
  {Steinberg}(2016)}]{Lucas16}%
  \BibitemOpen
  \bibfield  {author} {\bibinfo {author} {\bibfnamefont {Andrew}\ \bibnamefont
  {Lucas}}\ and\ \bibinfo {author} {\bibfnamefont {Julia}\ \bibnamefont
  {Steinberg}},\ }\bibfield  {title} {\enquote {\bibinfo {title} {Charge
  diffusion and the butterfly effect in striped holographic matter},}\ }\href
  {\doibase 10.1103/PhysRevLett.117.091601} {\bibfield  {journal} {\bibinfo
  {journal} {Journal of High Energy Physics}\ }\textbf {\bibinfo {volume}
  {10}},\ \bibinfo {pages} {142} (\bibinfo {year} {2016})}\BibitemShut
  {NoStop}%
\bibitem [{\citenamefont {Han}\ and\ \citenamefont {Hartnoll}(2019)}]{Han_vB}%
  \BibitemOpen
  \bibfield  {author} {\bibinfo {author} {\bibfnamefont {Xizhi}\ \bibnamefont
  {Han}}\ and\ \bibinfo {author} {\bibfnamefont {Sean~A.}\ \bibnamefont
  {Hartnoll}},\ }\bibfield  {title} {\enquote {\bibinfo {title} {{Quantum
  scrambling and state dependence of the butterfly velocity}},}\ }\href
  {\doibase 10.21468/SciPostPhys.7.4.045} {\bibfield  {journal} {\bibinfo
  {journal} {SciPost Phys.}\ }\textbf {\bibinfo {volume} {7}},\ \bibinfo
  {pages} {045} (\bibinfo {year} {2019})}\BibitemShut {NoStop}%
\bibitem [{\citenamefont {Yin}\ and\ \citenamefont {Lucas}(2022)}]{Yin_boson}%
  \BibitemOpen
  \bibfield  {author} {\bibinfo {author} {\bibfnamefont {Chao}\ \bibnamefont
  {Yin}}\ and\ \bibinfo {author} {\bibfnamefont {Andrew}\ \bibnamefont
  {Lucas}},\ }\bibfield  {title} {\enquote {\bibinfo {title} {Finite speed of
  quantum information in models of interacting bosons at finite density},}\
  }\href {\doibase 10.1103/PhysRevX.12.021039} {\bibfield  {journal} {\bibinfo
  {journal} {Phys. Rev. X}\ }\textbf {\bibinfo {volume} {12}},\ \bibinfo
  {pages} {021039} (\bibinfo {year} {2022})}\BibitemShut {NoStop}%
\bibitem [{\citenamefont {Lieb}\ and\ \citenamefont
  {Robinson}(1972)}]{Lieb1972}%
  \BibitemOpen
  \bibfield  {author} {\bibinfo {author} {\bibfnamefont {Elliott~H.}\
  \bibnamefont {Lieb}}\ and\ \bibinfo {author} {\bibfnamefont {Derek~W.}\
  \bibnamefont {Robinson}},\ }\bibfield  {title} {\enquote {\bibinfo {title}
  {The finite group velocity of quantum spin systems},}\ }\href {\doibase
  10.1007/BF01645779} {\bibfield  {journal} {\bibinfo  {journal} {Commun. Math.
  Phys.}\ }\textbf {\bibinfo {volume} {28}},\ \bibinfo {pages} {251--257}
  (\bibinfo {year} {1972})}\BibitemShut {NoStop}%
\bibitem [{\citenamefont {Blake}(2016{\natexlab{b}})}]{Blake_prd}%
  \BibitemOpen
  \bibfield  {author} {\bibinfo {author} {\bibfnamefont {Mike}\ \bibnamefont
  {Blake}},\ }\bibfield  {title} {\enquote {\bibinfo {title} {Universal
  diffusion in incoherent black holes},}\ }\href {\doibase
  10.1103/PhysRevD.94.086014} {\bibfield  {journal} {\bibinfo  {journal} {Phys.
  Rev. D}\ }\textbf {\bibinfo {volume} {94}},\ \bibinfo {pages} {086014}
  (\bibinfo {year} {2016}{\natexlab{b}})}\BibitemShut {NoStop}%
\bibitem [{\citenamefont {Patel}\ and\ \citenamefont {Sachdev}(2017)}]{E_diff}%
  \BibitemOpen
  \bibfield  {author} {\bibinfo {author} {\bibfnamefont {Aavishkar~A.}\
  \bibnamefont {Patel}}\ and\ \bibinfo {author} {\bibfnamefont {Subir}\
  \bibnamefont {Sachdev}},\ }\bibfield  {title} {\enquote {\bibinfo {title}
  {Quantum chaos on a critical fermi surface},}\ }\href {\doibase
  10.1073/pnas.1618185114} {\bibfield  {journal} {\bibinfo  {journal}
  {Proceedings of the National Academy of Sciences}\ }\textbf {\bibinfo
  {volume} {114}},\ \bibinfo {pages} {1844--1849} (\bibinfo {year}
  {2017})}\BibitemShut {NoStop}%
\bibitem [{\citenamefont {Xu}\ and\ \citenamefont
  {Swingle}(2022)}]{Swingle_rev22}%
  \BibitemOpen
  \bibfield  {author} {\bibinfo {author} {\bibfnamefont {Shenglong}\
  \bibnamefont {Xu}}\ and\ \bibinfo {author} {\bibfnamefont {Brian}\
  \bibnamefont {Swingle}},\ }\bibfield  {title} {\enquote {\bibinfo {title}
  {{Scrambling Dynamics and Out-of-Time Ordered Correlators in Quantum
  Many-Body Systems: a Tutorial}},}\ }\href@noop {} {\  (\bibinfo {year}
  {2022})},\ \Eprint {http://arxiv.org/abs/2202.07060} {arXiv:2202.07060
  [quant-ph]} \BibitemShut {NoStop}%
\bibitem [{\citenamefont {Huang}\ \emph {et~al.}(2017)\citenamefont {Huang},
  \citenamefont {Zhang},\ and\ \citenamefont {Chen}}]{ChenX16}%
  \BibitemOpen
  \bibfield  {author} {\bibinfo {author} {\bibfnamefont {Yichen}\ \bibnamefont
  {Huang}}, \bibinfo {author} {\bibfnamefont {Yong-Liang}\ \bibnamefont
  {Zhang}}, \ and\ \bibinfo {author} {\bibfnamefont {Xie}\ \bibnamefont
  {Chen}},\ }\bibfield  {title} {\enquote {\bibinfo {title}
  {Out-of-time-ordered correlators in many-body localized systems},}\ }\href
  {\doibase https://doi.org/10.1002/andp.201600318} {\bibfield  {journal}
  {\bibinfo  {journal} {Annalen der Physik}\ }\textbf {\bibinfo {volume}
  {529}},\ \bibinfo {pages} {1600318} (\bibinfo {year} {2017})}\BibitemShut
  {NoStop}%
\bibitem [{\citenamefont {Fan}\ \emph {et~al.}(2017)\citenamefont {Fan},
  \citenamefont {Zhang}, \citenamefont {Shen},\ and\ \citenamefont
  {Zhai}}]{Fan16}%
  \BibitemOpen
  \bibfield  {author} {\bibinfo {author} {\bibfnamefont {Ruihua}\ \bibnamefont
  {Fan}}, \bibinfo {author} {\bibfnamefont {Pengfei}\ \bibnamefont {Zhang}},
  \bibinfo {author} {\bibfnamefont {Huitao}\ \bibnamefont {Shen}}, \ and\
  \bibinfo {author} {\bibfnamefont {Hui}\ \bibnamefont {Zhai}},\ }\bibfield
  {title} {\enquote {\bibinfo {title} {Out-of-time-order correlation for
  many-body localization},}\ }\href {\doibase
  https://doi.org/10.1016/j.scib.2017.04.011} {\bibfield  {journal} {\bibinfo
  {journal} {Science Bulletin}\ }\textbf {\bibinfo {volume} {62}},\ \bibinfo
  {pages} {707--711} (\bibinfo {year} {2017})}\BibitemShut {NoStop}%
\bibitem [{\citenamefont {Chen}(2016)}]{Yu16}%
  \BibitemOpen
  \bibfield  {author} {\bibinfo {author} {\bibfnamefont {Yu}~\bibnamefont
  {Chen}},\ }\bibfield  {title} {\enquote {\bibinfo {title} {Universal
  logarithmic scrambling in many body localization},}\ }\href
  {https://arxiv.org/abs/1608.02765} {\bibfield  {journal} {\bibinfo  {journal}
  {arXiv: 1608.02765}\ } (\bibinfo {year} {2016})}\BibitemShut {NoStop}%
\bibitem [{\citenamefont {Swingle}\ and\ \citenamefont
  {Chowdhury}(2017)}]{Swingle16b}%
  \BibitemOpen
  \bibfield  {author} {\bibinfo {author} {\bibfnamefont {Brian}\ \bibnamefont
  {Swingle}}\ and\ \bibinfo {author} {\bibfnamefont {Debanjan}\ \bibnamefont
  {Chowdhury}},\ }\bibfield  {title} {\enquote {\bibinfo {title} {Slow
  scrambling in disordered quantum systems},}\ }\href {\doibase
  10.1103/PhysRevB.95.060201} {\bibfield  {journal} {\bibinfo  {journal} {Phys.
  Rev. B}\ }\textbf {\bibinfo {volume} {95}},\ \bibinfo {pages} {060201}
  (\bibinfo {year} {2017})}\BibitemShut {NoStop}%
\bibitem [{\citenamefont {He}\ and\ \citenamefont {Lu}(2017)}]{He16}%
  \BibitemOpen
  \bibfield  {author} {\bibinfo {author} {\bibfnamefont {Rong-Qiang}\
  \bibnamefont {He}}\ and\ \bibinfo {author} {\bibfnamefont {Zhong-Yi}\
  \bibnamefont {Lu}},\ }\bibfield  {title} {\enquote {\bibinfo {title}
  {Characterizing many-body localization by out-of-time-ordered correlation},}\
  }\href {\doibase 10.1103/PhysRevB.95.054201} {\bibfield  {journal} {\bibinfo
  {journal} {Phys. Rev. B}\ }\textbf {\bibinfo {volume} {95}},\ \bibinfo
  {pages} {054201} (\bibinfo {year} {2017})}\BibitemShut {NoStop}%
\bibitem [{\citenamefont {D\'ora}\ and\ \citenamefont
  {Moessner}(2017)}]{Luttinger17}%
  \BibitemOpen
  \bibfield  {author} {\bibinfo {author} {\bibfnamefont {Bal\'azs}\
  \bibnamefont {D\'ora}}\ and\ \bibinfo {author} {\bibfnamefont {Roderich}\
  \bibnamefont {Moessner}},\ }\bibfield  {title} {\enquote {\bibinfo {title}
  {Out-of-time-ordered density correlators in luttinger liquids},}\ }\href
  {\doibase 10.1103/PhysRevLett.119.026802} {\bibfield  {journal} {\bibinfo
  {journal} {Phys. Rev. Lett.}\ }\textbf {\bibinfo {volume} {119}},\ \bibinfo
  {pages} {026802} (\bibinfo {year} {2017})}\BibitemShut {NoStop}%
\bibitem [{\citenamefont {Bohrdt}\ \emph {et~al.}(2017)\citenamefont {Bohrdt},
  \citenamefont {Mendl}, \citenamefont {Endres},\ and\ \citenamefont
  {Knap}}]{Bohrdt17}%
  \BibitemOpen
  \bibfield  {author} {\bibinfo {author} {\bibfnamefont {A}~\bibnamefont
  {Bohrdt}}, \bibinfo {author} {\bibfnamefont {C~B}\ \bibnamefont {Mendl}},
  \bibinfo {author} {\bibfnamefont {M}~\bibnamefont {Endres}}, \ and\ \bibinfo
  {author} {\bibfnamefont {M}~\bibnamefont {Knap}},\ }\bibfield  {title}
  {\enquote {\bibinfo {title} {Scrambling and thermalization in a diffusive
  quantum many-body system},}\ }\href {\doibase 10.1088/1367-2630/aa719b}
  {\bibfield  {journal} {\bibinfo  {journal} {New Journal of Physics}\ }\textbf
  {\bibinfo {volume} {19}},\ \bibinfo {pages} {063001} (\bibinfo {year}
  {2017})}\BibitemShut {NoStop}%
\bibitem [{\citenamefont {Patel}\ \emph {et~al.}(2017)\citenamefont {Patel},
  \citenamefont {Chowdhury}, \citenamefont {Sachdev},\ and\ \citenamefont
  {Swingle}}]{Swingle17}%
  \BibitemOpen
  \bibfield  {author} {\bibinfo {author} {\bibfnamefont {Aavishkar~A.}\
  \bibnamefont {Patel}}, \bibinfo {author} {\bibfnamefont {Debanjan}\
  \bibnamefont {Chowdhury}}, \bibinfo {author} {\bibfnamefont {Subir}\
  \bibnamefont {Sachdev}}, \ and\ \bibinfo {author} {\bibfnamefont {Brian}\
  \bibnamefont {Swingle}},\ }\bibfield  {title} {\enquote {\bibinfo {title}
  {Quantum butterfly effect in weakly interacting diffusive metals},}\ }\href
  {\doibase 10.1103/PhysRevX.7.031047} {\bibfield  {journal} {\bibinfo
  {journal} {Phys. Rev. X}\ }\textbf {\bibinfo {volume} {7}},\ \bibinfo {pages}
  {031047} (\bibinfo {year} {2017})}\BibitemShut {NoStop}%
\bibitem [{\citenamefont {Liao}\ and\ \citenamefont {Galitski}(2018)}]{Liao18}%
  \BibitemOpen
  \bibfield  {author} {\bibinfo {author} {\bibfnamefont {Yunxiang}\
  \bibnamefont {Liao}}\ and\ \bibinfo {author} {\bibfnamefont {Victor}\
  \bibnamefont {Galitski}},\ }\bibfield  {title} {\enquote {\bibinfo {title}
  {Nonlinear sigma model approach to many-body quantum chaos: Regularized and
  unregularized out-of-time-ordered correlators},}\ }\href {\doibase
  10.1103/PhysRevB.98.205124} {\bibfield  {journal} {\bibinfo  {journal} {Phys.
  Rev. B}\ }\textbf {\bibinfo {volume} {98}},\ \bibinfo {pages} {205124}
  (\bibinfo {year} {2018})}\BibitemShut {NoStop}%
\bibitem [{\citenamefont {Li}\ \emph {et~al.}(2017)\citenamefont {Li},
  \citenamefont {Fan}, \citenamefont {Wang}, \citenamefont {Ye}, \citenamefont
  {Zeng}, \citenamefont {Zhai}, \citenamefont {Peng},\ and\ \citenamefont
  {Du}}]{Du17}%
  \BibitemOpen
  \bibfield  {author} {\bibinfo {author} {\bibfnamefont {Jun}\ \bibnamefont
  {Li}}, \bibinfo {author} {\bibfnamefont {Ruihua}\ \bibnamefont {Fan}},
  \bibinfo {author} {\bibfnamefont {Hengyan}\ \bibnamefont {Wang}}, \bibinfo
  {author} {\bibfnamefont {Bingtian}\ \bibnamefont {Ye}}, \bibinfo {author}
  {\bibfnamefont {Bei}\ \bibnamefont {Zeng}}, \bibinfo {author} {\bibfnamefont
  {Hui}\ \bibnamefont {Zhai}}, \bibinfo {author} {\bibfnamefont {Xinhua}\
  \bibnamefont {Peng}}, \ and\ \bibinfo {author} {\bibfnamefont {Jiangfeng}\
  \bibnamefont {Du}},\ }\bibfield  {title} {\enquote {\bibinfo {title}
  {Measuring out-of-time-order correlators on a nuclear magnetic resonance
  quantum simulator},}\ }\href {\doibase 10.1103/PhysRevX.7.031011} {\bibfield
  {journal} {\bibinfo  {journal} {Phys. Rev. X}\ }\textbf {\bibinfo {volume}
  {7}},\ \bibinfo {pages} {031011} (\bibinfo {year} {2017})}\BibitemShut
  {NoStop}%
\bibitem [{\citenamefont {Wei}\ \emph {et~al.}(2019)\citenamefont {Wei},
  \citenamefont {Peng}, \citenamefont {Shtanko}, \citenamefont {Marvian},
  \citenamefont {Lloyd}, \citenamefont {Ramanathan},\ and\ \citenamefont
  {Cappellaro}}]{pai_prl}%
  \BibitemOpen
  \bibfield  {author} {\bibinfo {author} {\bibfnamefont {Ken~Xuan}\
  \bibnamefont {Wei}}, \bibinfo {author} {\bibfnamefont {Pai}\ \bibnamefont
  {Peng}}, \bibinfo {author} {\bibfnamefont {Oles}\ \bibnamefont {Shtanko}},
  \bibinfo {author} {\bibfnamefont {Iman}\ \bibnamefont {Marvian}}, \bibinfo
  {author} {\bibfnamefont {Seth}\ \bibnamefont {Lloyd}}, \bibinfo {author}
  {\bibfnamefont {Chandrasekhar}\ \bibnamefont {Ramanathan}}, \ and\ \bibinfo
  {author} {\bibfnamefont {Paola}\ \bibnamefont {Cappellaro}},\ }\bibfield
  {title} {\enquote {\bibinfo {title} {Emergent prethermalization signatures in
  out-of-time ordered correlations},}\ }\href {\doibase
  10.1103/PhysRevLett.123.090605} {\bibfield  {journal} {\bibinfo  {journal}
  {Phys. Rev. Lett.}\ }\textbf {\bibinfo {volume} {123}},\ \bibinfo {pages}
  {090605} (\bibinfo {year} {2019})}\BibitemShut {NoStop}%
\bibitem [{\citenamefont {Nie}\ \emph {et~al.}(2020)\citenamefont {Nie},
  \citenamefont {Wei}, \citenamefont {Chen}, \citenamefont {Zhang},
  \citenamefont {Zhao}, \citenamefont {Qiu}, \citenamefont {Tian},
  \citenamefont {Ji}, \citenamefont {Xin}, \citenamefont {Lu},\ and\
  \citenamefont {Li}}]{Li20}%
  \BibitemOpen
  \bibfield  {author} {\bibinfo {author} {\bibfnamefont {Xinfang}\ \bibnamefont
  {Nie}}, \bibinfo {author} {\bibfnamefont {Bo-Bo}\ \bibnamefont {Wei}},
  \bibinfo {author} {\bibfnamefont {Xi}~\bibnamefont {Chen}}, \bibinfo {author}
  {\bibfnamefont {Ze}~\bibnamefont {Zhang}}, \bibinfo {author} {\bibfnamefont
  {Xiuzhu}\ \bibnamefont {Zhao}}, \bibinfo {author} {\bibfnamefont {Chudan}\
  \bibnamefont {Qiu}}, \bibinfo {author} {\bibfnamefont {Yu}~\bibnamefont
  {Tian}}, \bibinfo {author} {\bibfnamefont {Yunlan}\ \bibnamefont {Ji}},
  \bibinfo {author} {\bibfnamefont {Tao}\ \bibnamefont {Xin}}, \bibinfo
  {author} {\bibfnamefont {Dawei}\ \bibnamefont {Lu}}, \ and\ \bibinfo {author}
  {\bibfnamefont {Jun}\ \bibnamefont {Li}},\ }\bibfield  {title} {\enquote
  {\bibinfo {title} {Experimental observation of equilibrium and dynamical
  quantum phase transitions via out-of-time-ordered correlators},}\ }\href
  {\doibase 10.1103/PhysRevLett.124.250601} {\bibfield  {journal} {\bibinfo
  {journal} {Phys. Rev. Lett.}\ }\textbf {\bibinfo {volume} {124}},\ \bibinfo
  {pages} {250601} (\bibinfo {year} {2020})}\BibitemShut {NoStop}%
\bibitem [{\citenamefont {G\"arttner}\ \emph {et~al.}(2017)\citenamefont
  {G\"arttner}, \citenamefont {Bohnet}, \citenamefont {Safavi-Naini},
  \citenamefont {Wall}, \citenamefont {Bollinger},\ and\ \citenamefont
  {Rey}}]{otoc_bolinger}%
  \BibitemOpen
  \bibfield  {author} {\bibinfo {author} {\bibfnamefont {Martin}\ \bibnamefont
  {G\"arttner}}, \bibinfo {author} {\bibfnamefont {Justin~G.}\ \bibnamefont
  {Bohnet}}, \bibinfo {author} {\bibfnamefont {Arghavan}\ \bibnamefont
  {Safavi-Naini}}, \bibinfo {author} {\bibfnamefont {Michael~L.}\ \bibnamefont
  {Wall}}, \bibinfo {author} {\bibfnamefont {John~J.}\ \bibnamefont
  {Bollinger}}, \ and\ \bibinfo {author} {\bibfnamefont {Ana~Maria}\
  \bibnamefont {Rey}},\ }\bibfield  {title} {\enquote {\bibinfo {title}
  {{Measuring out-of-time-order correlations and multiple quantum spectra in a
  trapped ion quantum magnet}},}\ }\href {\doibase 10.1038/nphys4119}
  {\bibfield  {journal} {\bibinfo  {journal} {Nature Phys.}\ }\textbf {\bibinfo
  {volume} {13}},\ \bibinfo {pages} {781} (\bibinfo {year} {2017})},\ \Eprint
  {http://arxiv.org/abs/1608.08938} {arXiv:1608.08938 [quant-ph]} \BibitemShut
  {NoStop}%
\bibitem [{\citenamefont {Joshi}\ \emph {et~al.}(2020)\citenamefont {Joshi},
  \citenamefont {Elben}, \citenamefont {Vermersch}, \citenamefont {Brydges},
  \citenamefont {Maier}, \citenamefont {Zoller}, \citenamefont {Blatt},\ and\
  \citenamefont {Roos}}]{IonTrap20}%
  \BibitemOpen
  \bibfield  {author} {\bibinfo {author} {\bibfnamefont {Manoj~K.}\
  \bibnamefont {Joshi}}, \bibinfo {author} {\bibfnamefont {Andreas}\
  \bibnamefont {Elben}}, \bibinfo {author} {\bibfnamefont {Beno\^{\i}t}\
  \bibnamefont {Vermersch}}, \bibinfo {author} {\bibfnamefont {Tiff}\
  \bibnamefont {Brydges}}, \bibinfo {author} {\bibfnamefont {Christine}\
  \bibnamefont {Maier}}, \bibinfo {author} {\bibfnamefont {Peter}\ \bibnamefont
  {Zoller}}, \bibinfo {author} {\bibfnamefont {Rainer}\ \bibnamefont {Blatt}},
  \ and\ \bibinfo {author} {\bibfnamefont {Christian~F.}\ \bibnamefont
  {Roos}},\ }\bibfield  {title} {\enquote {\bibinfo {title} {Quantum
  information scrambling in a trapped-ion quantum simulator with tunable range
  interactions},}\ }\href {\doibase 10.1103/PhysRevLett.124.240505} {\bibfield
  {journal} {\bibinfo  {journal} {Phys. Rev. Lett.}\ }\textbf {\bibinfo
  {volume} {124}},\ \bibinfo {pages} {240505} (\bibinfo {year}
  {2020})}\BibitemShut {NoStop}%
\bibitem [{\citenamefont {Blok}\ \emph {et~al.}(2021)\citenamefont {Blok},
  \citenamefont {Ramasesh}, \citenamefont {Schuster}, \citenamefont {O'Brien},
  \citenamefont {Kreikebaum}, \citenamefont {Dahlen}, \citenamefont {Morvan},
  \citenamefont {Yoshida}, \citenamefont {Yao},\ and\ \citenamefont
  {Siddiqi}}]{Yao21}%
  \BibitemOpen
  \bibfield  {author} {\bibinfo {author} {\bibfnamefont {M.~S.}\ \bibnamefont
  {Blok}}, \bibinfo {author} {\bibfnamefont {V.~V.}\ \bibnamefont {Ramasesh}},
  \bibinfo {author} {\bibfnamefont {T.}~\bibnamefont {Schuster}}, \bibinfo
  {author} {\bibfnamefont {K.}~\bibnamefont {O'Brien}}, \bibinfo {author}
  {\bibfnamefont {J.~M.}\ \bibnamefont {Kreikebaum}}, \bibinfo {author}
  {\bibfnamefont {D.}~\bibnamefont {Dahlen}}, \bibinfo {author} {\bibfnamefont
  {A.}~\bibnamefont {Morvan}}, \bibinfo {author} {\bibfnamefont
  {B.}~\bibnamefont {Yoshida}}, \bibinfo {author} {\bibfnamefont {N.~Y.}\
  \bibnamefont {Yao}}, \ and\ \bibinfo {author} {\bibfnamefont
  {I.}~\bibnamefont {Siddiqi}},\ }\bibfield  {title} {\enquote {\bibinfo
  {title} {Quantum information scrambling on a superconducting qutrit
  processor},}\ }\href {\doibase 10.1103/PhysRevX.11.021010} {\bibfield
  {journal} {\bibinfo  {journal} {Phys. Rev. X}\ }\textbf {\bibinfo {volume}
  {11}},\ \bibinfo {pages} {021010} (\bibinfo {year} {2021})}\BibitemShut
  {NoStop}%
\bibitem [{\citenamefont {Mi}\ \emph {et~al.}(2021)\citenamefont {Mi},
  \citenamefont {Roushan}, \citenamefont {Quintana}, \citenamefont {Mandra},
  \citenamefont {Marshall}, \citenamefont {Neill}, \citenamefont {Arute},
  \citenamefont {Arya}, \citenamefont {Atalaya}, \citenamefont {Babbush} \emph
  {et~al.}}]{Google21}%
  \BibitemOpen
  \bibfield  {author} {\bibinfo {author} {\bibfnamefont {Xiao}\ \bibnamefont
  {Mi}}, \bibinfo {author} {\bibfnamefont {Pedram}\ \bibnamefont {Roushan}},
  \bibinfo {author} {\bibfnamefont {Chris}\ \bibnamefont {Quintana}}, \bibinfo
  {author} {\bibfnamefont {Salvatore}\ \bibnamefont {Mandra}}, \bibinfo
  {author} {\bibfnamefont {Jeffrey}\ \bibnamefont {Marshall}}, \bibinfo
  {author} {\bibfnamefont {Charles}\ \bibnamefont {Neill}}, \bibinfo {author}
  {\bibfnamefont {Frank}\ \bibnamefont {Arute}}, \bibinfo {author}
  {\bibfnamefont {Kunal}\ \bibnamefont {Arya}}, \bibinfo {author}
  {\bibfnamefont {Juan}\ \bibnamefont {Atalaya}}, \bibinfo {author}
  {\bibfnamefont {Ryan}\ \bibnamefont {Babbush}},  \emph {et~al.},\ }\bibfield
  {title} {\enquote {\bibinfo {title} {Information scrambling in quantum
  circuits},}\ }\href {\doibase 10.1126/science.abg5029} {\bibfield  {journal}
  {\bibinfo  {journal} {Science}\ }\textbf {\bibinfo {volume} {374}},\ \bibinfo
  {pages} {1479--1483} (\bibinfo {year} {2021})}\BibitemShut {NoStop}%
\bibitem [{\citenamefont {Braum{\"u}ller}\ \emph {et~al.}(2022)\citenamefont
  {Braum{\"u}ller}, \citenamefont {Karamlou}, \citenamefont {Yanay},
  \citenamefont {Kannan}, \citenamefont {Kim}, \citenamefont {Kjaergaard},
  \citenamefont {Melville}, \citenamefont {Niedzielski}, \citenamefont {Sung},
  \citenamefont {Veps{\"a}l{\"a}inen} \emph {et~al.}}]{otoc_sc22}%
  \BibitemOpen
  \bibfield  {author} {\bibinfo {author} {\bibfnamefont {Jochen}\ \bibnamefont
  {Braum{\"u}ller}}, \bibinfo {author} {\bibfnamefont {Amir~H}\ \bibnamefont
  {Karamlou}}, \bibinfo {author} {\bibfnamefont {Yariv}\ \bibnamefont {Yanay}},
  \bibinfo {author} {\bibfnamefont {Bharath}\ \bibnamefont {Kannan}}, \bibinfo
  {author} {\bibfnamefont {David}\ \bibnamefont {Kim}}, \bibinfo {author}
  {\bibfnamefont {Morten}\ \bibnamefont {Kjaergaard}}, \bibinfo {author}
  {\bibfnamefont {Alexander}\ \bibnamefont {Melville}}, \bibinfo {author}
  {\bibfnamefont {Bethany~M}\ \bibnamefont {Niedzielski}}, \bibinfo {author}
  {\bibfnamefont {Youngkyu}\ \bibnamefont {Sung}}, \bibinfo {author}
  {\bibfnamefont {Antti}\ \bibnamefont {Veps{\"a}l{\"a}inen}},  \emph
  {et~al.},\ }\bibfield  {title} {\enquote {\bibinfo {title} {Probing quantum
  information propagation with out-of-time-ordered correlators},}\ }\href
  {\doibase https://doi.org/10.1038/s41567-021-01430-w} {\bibfield  {journal}
  {\bibinfo  {journal} {Nature Physics}\ }\textbf {\bibinfo {volume} {18}},\
  \bibinfo {pages} {172--178} (\bibinfo {year} {2022})}\BibitemShut {NoStop}%
\bibitem [{\citenamefont {Anderson}\ \emph {et~al.}(1995)\citenamefont
  {Anderson}, \citenamefont {Ensher}, \citenamefont {Matthews}, \citenamefont
  {Wieman},\ and\ \citenamefont {Cornell}}]{BEC95}%
  \BibitemOpen
  \bibfield  {author} {\bibinfo {author} {\bibfnamefont {M.~H.}\ \bibnamefont
  {Anderson}}, \bibinfo {author} {\bibfnamefont {J.~R.}\ \bibnamefont
  {Ensher}}, \bibinfo {author} {\bibfnamefont {M.~R.}\ \bibnamefont
  {Matthews}}, \bibinfo {author} {\bibfnamefont {C.~E.}\ \bibnamefont
  {Wieman}}, \ and\ \bibinfo {author} {\bibfnamefont {E.~A.}\ \bibnamefont
  {Cornell}},\ }\bibfield  {title} {\enquote {\bibinfo {title} {Observation of
  bose-einstein condensation in a dilute atomic vapor},}\ }\href {\doibase
  10.1126/science.269.5221.198} {\bibfield  {journal} {\bibinfo  {journal}
  {Science}\ }\textbf {\bibinfo {volume} {269}},\ \bibinfo {pages} {198--201}
  (\bibinfo {year} {1995})}\BibitemShut {NoStop}%
\bibitem [{\citenamefont {Aleiner}\ \emph {et~al.}(2016)\citenamefont
  {Aleiner}, \citenamefont {Faoro},\ and\ \citenamefont {Ioffe}}]{Aleiner16}%
  \BibitemOpen
  \bibfield  {author} {\bibinfo {author} {\bibfnamefont {Igor~L.}\ \bibnamefont
  {Aleiner}}, \bibinfo {author} {\bibfnamefont {Lara}\ \bibnamefont {Faoro}}, \
  and\ \bibinfo {author} {\bibfnamefont {Lev~B.}\ \bibnamefont {Ioffe}},\
  }\bibfield  {title} {\enquote {\bibinfo {title} {Microscopic model of quantum
  butterfly effect: Out-of-time-order correlators and traveling combustion
  waves},}\ }\href {\doibase https://doi.org/10.1016/j.aop.2016.09.006}
  {\bibfield  {journal} {\bibinfo  {journal} {Annals of Physics}\ }\textbf
  {\bibinfo {volume} {375}},\ \bibinfo {pages} {378--406} (\bibinfo {year}
  {2016})}\BibitemShut {NoStop}%
\bibitem [{\citenamefont {Klug}\ \emph {et~al.}(2018)\citenamefont {Klug},
  \citenamefont {Scheurer},\ and\ \citenamefont {Schmalian}}]{graphene_otoc}%
  \BibitemOpen
  \bibfield  {author} {\bibinfo {author} {\bibfnamefont {Markus~J.}\
  \bibnamefont {Klug}}, \bibinfo {author} {\bibfnamefont {Mathias~S.}\
  \bibnamefont {Scheurer}}, \ and\ \bibinfo {author} {\bibfnamefont {J\"org}\
  \bibnamefont {Schmalian}},\ }\bibfield  {title} {\enquote {\bibinfo {title}
  {Hierarchy of information scrambling, thermalization, and hydrodynamic flow
  in graphene},}\ }\href {\doibase 10.1103/PhysRevB.98.045102} {\bibfield
  {journal} {\bibinfo  {journal} {Phys. Rev. B}\ }\textbf {\bibinfo {volume}
  {98}},\ \bibinfo {pages} {045102} (\bibinfo {year} {2018})}\BibitemShut
  {NoStop}%
\bibitem [{\citenamefont {Grozdanov}\ \emph {et~al.}(2019)\citenamefont
  {Grozdanov}, \citenamefont {Schalm},\ and\ \citenamefont
  {Scopelliti}}]{gBE_PRE}%
  \BibitemOpen
  \bibfield  {author} {\bibinfo {author} {\bibfnamefont {Sa\ifmmode
  \check{s}\else~\v{s}\fi{}o}\ \bibnamefont {Grozdanov}}, \bibinfo {author}
  {\bibfnamefont {Koenraad}\ \bibnamefont {Schalm}}, \ and\ \bibinfo {author}
  {\bibfnamefont {Vincenzo}\ \bibnamefont {Scopelliti}},\ }\bibfield  {title}
  {\enquote {\bibinfo {title} {Kinetic theory for classical and quantum
  many-body chaos},}\ }\href {\doibase 10.1103/PhysRevE.99.012206} {\bibfield
  {journal} {\bibinfo  {journal} {Phys. Rev. E}\ }\textbf {\bibinfo {volume}
  {99}},\ \bibinfo {pages} {012206} (\bibinfo {year} {2019})}\BibitemShut
  {NoStop}%
\bibitem [{\citenamefont {Zhang}(2019)}]{Pengfei19}%
  \BibitemOpen
  \bibfield  {author} {\bibinfo {author} {\bibfnamefont {Pengfei}\ \bibnamefont
  {Zhang}},\ }\bibfield  {title} {\enquote {\bibinfo {title} {Quantum chaos for
  the unitary fermi gas from the generalized boltzmann equations},}\ }\href
  {\doibase 10.1088/1361-6455/ab0af9} {\bibfield  {journal} {\bibinfo
  {journal} {Journal of Physics B: Atomic, Molecular and Optical Physics}\
  }\textbf {\bibinfo {volume} {52}},\ \bibinfo {pages} {135301} (\bibinfo
  {year} {2019})}\BibitemShut {NoStop}%
\bibitem [{\citenamefont {Zhai}(2021)}]{zhai_book}%
  \BibitemOpen
  \bibfield  {author} {\bibinfo {author} {\bibfnamefont {H.}~\bibnamefont
  {Zhai}},\ }\href {https://books.google.com/books?id=5jcTEAAAQBAJ} {\emph
  {\bibinfo {title} {Ultracold Atomic Physics}}}\ (\bibinfo  {publisher}
  {Cambridge University Press},\ \bibinfo {year} {2021})\BibitemShut {NoStop}%
\bibitem [{\citenamefont {Imamovic-Tomasovic}\ and\ \citenamefont
  {Griffin}(2001)}]{Matrix01}%
  \BibitemOpen
  \bibfield  {author} {\bibinfo {author} {\bibfnamefont {M}~\bibnamefont
  {Imamovic-Tomasovic}}\ and\ \bibinfo {author} {\bibfnamefont {A}~\bibnamefont
  {Griffin}},\ }\bibfield  {title} {\enquote {\bibinfo {title} {Quasiparticle
  kinetic equation in a trapped bose gas at low temperatures},}\ }\href
  {\doibase https://doi.org/10.1023/A:1004860602930} {\bibfield  {journal}
  {\bibinfo  {journal} {Journal of low temperature physics}\ }\textbf {\bibinfo
  {volume} {122}},\ \bibinfo {pages} {617--655} (\bibinfo {year}
  {2001})}\BibitemShut {NoStop}%
\bibitem [{\citenamefont {Romero-Berm\'udez}\ \emph {et~al.}(2019)\citenamefont
  {Romero-Berm\'udez}, \citenamefont {Schalm},\ and\ \citenamefont
  {Scopelliti}}]{regularization19}%
  \BibitemOpen
  \bibfield  {author} {\bibinfo {author} {\bibfnamefont {Aurelio}\ \bibnamefont
  {Romero-Berm\'udez}}, \bibinfo {author} {\bibfnamefont {Koenraad}\
  \bibnamefont {Schalm}}, \ and\ \bibinfo {author} {\bibfnamefont {Vincenzo}\
  \bibnamefont {Scopelliti}},\ }\bibfield  {title} {\enquote {\bibinfo {title}
  {{Regularization dependence of the OTOC. Which Lyapunov spectrum is the
  physical one?}}}\ }\href {\doibase 10.1007/JHEP07(2019)107} {\bibfield
  {journal} {\bibinfo  {journal} {JHEP}\ }\textbf {\bibinfo {volume} {07}},\
  \bibinfo {pages} {107} (\bibinfo {year} {2019})},\ \Eprint
  {http://arxiv.org/abs/1903.09595} {arXiv:1903.09595 [hep-th]} \BibitemShut
  {NoStop}%
\bibitem [{\citenamefont {Kamenev}(2011)}]{kamenev_book}%
  \BibitemOpen
  \bibfield  {author} {\bibinfo {author} {\bibfnamefont {A.}~\bibnamefont
  {Kamenev}},\ }\href {https://books.google.com/books?id=CwlrUepnla4C} {\emph
  {\bibinfo {title} {Field Theory of Non-Equilibrium Systems}}}\ (\bibinfo
  {publisher} {Cambridge University Press},\ \bibinfo {year}
  {2011})\BibitemShut {NoStop}%
\bibitem [{\citenamefont {Beliaev}(1958)}]{beliaev58}%
  \BibitemOpen
  \bibfield  {author} {\bibinfo {author} {\bibfnamefont {ST}~\bibnamefont
  {Beliaev}},\ }\bibfield  {title} {\enquote {\bibinfo {title} {Energy spectrum
  of a non-ideal bose gas},}\ }\href@noop {} {\bibfield  {journal} {\bibinfo
  {journal} {Sov. Phys. JETP}\ }\textbf {\bibinfo {volume} {7}},\ \bibinfo
  {pages} {299--307} (\bibinfo {year} {1958})}\BibitemShut {NoStop}%
\bibitem [{Note1()}]{Note1}%
  \BibitemOpen
  \bibinfo {note} {Since $\delta F_{k}^{du}$ represents a density in $\protect
  \bm {k}$ space and the system is isotropic, $k^2\delta F^{du}_k$ represents
  the density of $k$ with all angular directions being integrated.}\BibitemShut
  {Stop}%
\bibitem [{\citenamefont {Szépfalusy}\ and\ \citenamefont
  {Kondor}(1974)}]{LandauD_74}%
  \BibitemOpen
  \bibfield  {author} {\bibinfo {author} {\bibfnamefont {P}~\bibnamefont
  {Szépfalusy}}\ and\ \bibinfo {author} {\bibfnamefont {I}~\bibnamefont
  {Kondor}},\ }\bibfield  {title} {\enquote {\bibinfo {title} {On the dynamics
  of continuous phase transitions},}\ }\href {\doibase
  https://doi.org/10.1016/0003-4916(74)90330-3} {\bibfield  {journal} {\bibinfo
   {journal} {Annals of Physics}\ }\textbf {\bibinfo {volume} {82}},\ \bibinfo
  {pages} {1--53} (\bibinfo {year} {1974})}\BibitemShut {NoStop}%
\bibitem [{\citenamefont {Chung}\ and\ \citenamefont
  {Bhattacherjee}(2009)}]{damp09}%
  \BibitemOpen
  \bibfield  {author} {\bibinfo {author} {\bibfnamefont {Ming-Chiang}\
  \bibnamefont {Chung}}\ and\ \bibinfo {author} {\bibfnamefont {Aranya~B}\
  \bibnamefont {Bhattacherjee}},\ }\bibfield  {title} {\enquote {\bibinfo
  {title} {Damping in 2d and 3d dilute bose gases},}\ }\href {\doibase
  10.1088/1367-2630/11/12/123012} {\bibfield  {journal} {\bibinfo  {journal}
  {New Journal of Physics}\ }\textbf {\bibinfo {volume} {11}},\ \bibinfo
  {pages} {123012} (\bibinfo {year} {2009})}\BibitemShut {NoStop}%
\bibitem [{\citenamefont {Bilitewski}\ \emph {et~al.}(2018)\citenamefont
  {Bilitewski}, \citenamefont {Bhattacharjee},\ and\ \citenamefont
  {Moessner}}]{classical_chain_vb}%
  \BibitemOpen
  \bibfield  {author} {\bibinfo {author} {\bibfnamefont {Thomas}\ \bibnamefont
  {Bilitewski}}, \bibinfo {author} {\bibfnamefont {Subhro}\ \bibnamefont
  {Bhattacharjee}}, \ and\ \bibinfo {author} {\bibfnamefont {Roderich}\
  \bibnamefont {Moessner}},\ }\bibfield  {title} {\enquote {\bibinfo {title}
  {Temperature dependence of the butterfly effect in a classical many-body
  system},}\ }\href {\doibase 10.1103/PhysRevLett.121.250602} {\bibfield
  {journal} {\bibinfo  {journal} {Phys. Rev. Lett.}\ }\textbf {\bibinfo
  {volume} {121}},\ \bibinfo {pages} {250602} (\bibinfo {year}
  {2018})}\BibitemShut {NoStop}%
\bibitem [{\citenamefont {Gu}\ \emph {et~al.}(2017{\natexlab{a}})\citenamefont
  {Gu}, \citenamefont {Qi},\ and\ \citenamefont {Stanford}}]{YIngfei17a}%
  \BibitemOpen
  \bibfield  {author} {\bibinfo {author} {\bibfnamefont {Yingfei}\ \bibnamefont
  {Gu}}, \bibinfo {author} {\bibfnamefont {Xiao-Liang}\ \bibnamefont {Qi}}, \
  and\ \bibinfo {author} {\bibfnamefont {Douglas}\ \bibnamefont {Stanford}},\
  }\bibfield  {title} {\enquote {\bibinfo {title} {Local criticality, diffusion
  and chaos in generalized sachdev-ye-kitaev models},}\ }\href {\doibase
  10.1073/pnas.1618185114} {\bibfield  {journal} {\bibinfo  {journal} {Journal
  of High Energy Physics}\ }\textbf {\bibinfo {volume} {05}},\ \bibinfo {pages}
  {125} (\bibinfo {year} {2017}{\natexlab{a}})}\BibitemShut {NoStop}%
\bibitem [{\citenamefont {Davison}\ \emph {et~al.}(2017)\citenamefont
  {Davison}, \citenamefont {Fu}, \citenamefont {Georges}, \citenamefont {Gu},
  \citenamefont {Jensen},\ and\ \citenamefont {Sachdev}}]{Sachdev17b}%
  \BibitemOpen
  \bibfield  {author} {\bibinfo {author} {\bibfnamefont {Richard~A.}\
  \bibnamefont {Davison}}, \bibinfo {author} {\bibfnamefont {Wenbo}\
  \bibnamefont {Fu}}, \bibinfo {author} {\bibfnamefont {Antoine}\ \bibnamefont
  {Georges}}, \bibinfo {author} {\bibfnamefont {Yingfei}\ \bibnamefont {Gu}},
  \bibinfo {author} {\bibfnamefont {Kristan}\ \bibnamefont {Jensen}}, \ and\
  \bibinfo {author} {\bibfnamefont {Subir}\ \bibnamefont {Sachdev}},\
  }\bibfield  {title} {\enquote {\bibinfo {title} {Thermoelectric transport in
  disordered metals without quasiparticles: The sachdev-ye-kitaev models and
  holography},}\ }\href {\doibase 10.1103/PhysRevB.95.155131} {\bibfield
  {journal} {\bibinfo  {journal} {Phys. Rev. B}\ }\textbf {\bibinfo {volume}
  {95}},\ \bibinfo {pages} {155131} (\bibinfo {year} {2017})}\BibitemShut
  {NoStop}%
\bibitem [{\citenamefont {Gu}\ \emph {et~al.}(2017{\natexlab{b}})\citenamefont
  {Gu}, \citenamefont {Lucas},\ and\ \citenamefont {Qi}}]{Lucas17}%
  \BibitemOpen
  \bibfield  {author} {\bibinfo {author} {\bibfnamefont {Yingfei}\ \bibnamefont
  {Gu}}, \bibinfo {author} {\bibfnamefont {Andrew}\ \bibnamefont {Lucas}}, \
  and\ \bibinfo {author} {\bibfnamefont {Xiao-Liang}\ \bibnamefont {Qi}},\
  }\bibfield  {title} {\enquote {\bibinfo {title} {{Energy diffusion and the
  butterfly effect in inhomogeneous Sachdev-Ye-Kitaev chains}},}\ }\href
  {\doibase 10.21468/SciPostPhys.2.3.018} {\bibfield  {journal} {\bibinfo
  {journal} {SciPost Phys.}\ }\textbf {\bibinfo {volume} {2}},\ \bibinfo
  {pages} {018} (\bibinfo {year} {2017}{\natexlab{b}})}\BibitemShut {NoStop}%
\bibitem [{\citenamefont {Werman}\ \emph {et~al.}(2017)\citenamefont {Werman},
  \citenamefont {Kivelson},\ and\ \citenamefont {Berg}}]{otoc_ele_phon}%
  \BibitemOpen
  \bibfield  {author} {\bibinfo {author} {\bibfnamefont {Yochai}\ \bibnamefont
  {Werman}}, \bibinfo {author} {\bibfnamefont {Steven~A}\ \bibnamefont
  {Kivelson}}, \ and\ \bibinfo {author} {\bibfnamefont {Erez}\ \bibnamefont
  {Berg}},\ }\bibfield  {title} {\enquote {\bibinfo {title} {Quantum chaos in
  an electron-phonon bad metal},}\ }\href {https://arxiv.org/abs/1705.07895}
  {\bibfield  {journal} {\bibinfo  {journal} {arXiv:1705.07895}\ } (\bibinfo
  {year} {2017})}\BibitemShut {NoStop}%
\bibitem [{\citenamefont {Kirkpatrick}\ and\ \citenamefont
  {Dorfman}(1983)}]{kirk83}%
  \BibitemOpen
  \bibfield  {author} {\bibinfo {author} {\bibfnamefont {T.~R.}\ \bibnamefont
  {Kirkpatrick}}\ and\ \bibinfo {author} {\bibfnamefont {J.~R.}\ \bibnamefont
  {Dorfman}},\ }\bibfield  {title} {\enquote {\bibinfo {title} {Transport
  theory for a weakly interacting condensed bose gas},}\ }\href {\doibase
  10.1103/PhysRevA.28.2576} {\bibfield  {journal} {\bibinfo  {journal} {Phys.
  Rev. A}\ }\textbf {\bibinfo {volume} {28}},\ \bibinfo {pages} {2576--2579}
  (\bibinfo {year} {1983})}\BibitemShut {NoStop}%
\bibitem [{\citenamefont {Kirkpatrick}\ and\ \citenamefont
  {Dorfman}(1985)}]{kirk85}%
  \BibitemOpen
  \bibfield  {author} {\bibinfo {author} {\bibfnamefont {TR}~\bibnamefont
  {Kirkpatrick}}\ and\ \bibinfo {author} {\bibfnamefont {JR}~\bibnamefont
  {Dorfman}},\ }\bibfield  {title} {\enquote {\bibinfo {title} {Transport
  coefficients in a dilute but condensed bose gas},}\ }\href {\doibase
  https://doi.org/10.1007/BF00681133} {\bibfield  {journal} {\bibinfo
  {journal} {Journal of low temperature physics}\ }\textbf {\bibinfo {volume}
  {58}},\ \bibinfo {pages} {399--415} (\bibinfo {year} {1985})}\BibitemShut
  {NoStop}%
\bibitem [{\citenamefont {Pethick}\ and\ \citenamefont
  {Smith}(2008)}]{BEC_book}%
  \BibitemOpen
  \bibfield  {author} {\bibinfo {author} {\bibfnamefont {C.J.}\ \bibnamefont
  {Pethick}}\ and\ \bibinfo {author} {\bibfnamefont {H.}~\bibnamefont
  {Smith}},\ }\href {https://books.google.com/books?id=G8kgAwAAQBAJ} {\emph
  {\bibinfo {title} {Bose--Einstein Condensation in Dilute Gases}}}\ (\bibinfo
  {publisher} {Cambridge University Press},\ \bibinfo {year}
  {2008})\BibitemShut {NoStop}%
\bibitem [{\citenamefont {Swingle}\ \emph {et~al.}(2016)\citenamefont
  {Swingle}, \citenamefont {Bentsen}, \citenamefont {Schleier-Smith},\ and\
  \citenamefont {Hayden}}]{measure_otoc16}%
  \BibitemOpen
  \bibfield  {author} {\bibinfo {author} {\bibfnamefont {Brian}\ \bibnamefont
  {Swingle}}, \bibinfo {author} {\bibfnamefont {Gregory}\ \bibnamefont
  {Bentsen}}, \bibinfo {author} {\bibfnamefont {Monika}\ \bibnamefont
  {Schleier-Smith}}, \ and\ \bibinfo {author} {\bibfnamefont {Patrick}\
  \bibnamefont {Hayden}},\ }\bibfield  {title} {\enquote {\bibinfo {title}
  {Measuring the scrambling of quantum information},}\ }\href {\doibase
  10.1103/PhysRevA.94.040302} {\bibfield  {journal} {\bibinfo  {journal} {Phys.
  Rev. A}\ }\textbf {\bibinfo {volume} {94}},\ \bibinfo {pages} {040302}
  (\bibinfo {year} {2016})}\BibitemShut {NoStop}%
\bibitem [{\citenamefont {Vermersch}\ \emph {et~al.}(2019)\citenamefont
  {Vermersch}, \citenamefont {Elben}, \citenamefont {Sieberer}, \citenamefont
  {Yao},\ and\ \citenamefont {Zoller}}]{measure_otoc19}%
  \BibitemOpen
  \bibfield  {author} {\bibinfo {author} {\bibfnamefont {B.}~\bibnamefont
  {Vermersch}}, \bibinfo {author} {\bibfnamefont {A.}~\bibnamefont {Elben}},
  \bibinfo {author} {\bibfnamefont {L.~M.}\ \bibnamefont {Sieberer}}, \bibinfo
  {author} {\bibfnamefont {N.~Y.}\ \bibnamefont {Yao}}, \ and\ \bibinfo
  {author} {\bibfnamefont {P.}~\bibnamefont {Zoller}},\ }\bibfield  {title}
  {\enquote {\bibinfo {title} {Probing scrambling using statistical
  correlations between randomized measurements},}\ }\href {\doibase
  10.1103/PhysRevX.9.021061} {\bibfield  {journal} {\bibinfo  {journal} {Phys.
  Rev. X}\ }\textbf {\bibinfo {volume} {9}},\ \bibinfo {pages} {021061}
  (\bibinfo {year} {2019})}\BibitemShut {NoStop}%
\bibitem [{\citenamefont {Griffin}\ \emph {et~al.}(2009)\citenamefont
  {Griffin}, \citenamefont {Nikuni},\ and\ \citenamefont
  {Zaremba}}]{BEC_T_book}%
  \BibitemOpen
  \bibfield  {author} {\bibinfo {author} {\bibfnamefont {A.}~\bibnamefont
  {Griffin}}, \bibinfo {author} {\bibfnamefont {T.}~\bibnamefont {Nikuni}}, \
  and\ \bibinfo {author} {\bibfnamefont {E.}~\bibnamefont {Zaremba}},\ }\href
  {https://books.google.com/books?id=yjDRQ\_e5UZIC} {\emph {\bibinfo {title}
  {Bose-Condensed Gases at Finite Temperatures}}}\ (\bibinfo  {publisher}
  {Cambridge University Press},\ \bibinfo {year} {2009})\BibitemShut {NoStop}%
\end{thebibliography}%

\end{document}